\documentclass[aps,prd,12pt,superscriptaddress,showpacs,showkeys,reprint,nofootinbib]{revtex4-1}
\usepackage{calrsfs} 
\usepackage{amsmath,amsthm,latexsym,amssymb,amsfonts}
\usepackage{xcolor}
\usepackage{textcomp}
\usepackage{float}
\usepackage[%
  colorlinks=true,
  urlcolor=blue,
  linkcolor=blue,
  citecolor=blue
]{hyperref}
\usepackage{etoolbox}
\usepackage{breqn}
\allowdisplaybreaks 

\usepackage{mathtools,amsmath,amssymb,amsfonts,dsfont,mathrsfs,amsthm,bm,latexsym}
\usepackage{graphicx}
\usepackage{centernot}
\usepackage{xcolor}
\usepackage{comment}
\usepackage{feynmf}
\usepackage{siunitx}
\usepackage{array}
\newcolumntype{L}[1]{>{\raggedright\let\newline\\\arraybackslash\hspace{0pt}}m{#1}}
\newcolumntype{C}[1]{>{\centering\let\newline\\\arraybackslash\hspace{0pt}}m{#1}}
\newcolumntype{R}[1]{>{\raggedleft\let\newline\\\arraybackslash\hspace{0pt}}m{#1}}
\usepackage{braket}
\usepackage{xparse}

\DeclareDocumentCommand{\Ag}{ s o }{ \IfBooleanTF{#1}
    { \IfValueTF{#2}{ \bm{\mathcal{A}}_{(#2)} }{ \bm{\mathcal{A}} } }
    { \IfValueTF{#2}{    {\mathcal{A}}_{(#2)} }{    {\mathcal{A}} } } }
\DeclareDocumentCommand{\Af}{ s o }{ \IfBooleanTF{#1}
    { \IfValueTF{#2}{ \boldsymbol{A}_{(#2)} }{ \boldsymbol{A} } }
    { \IfValueTF{#2}{            {A}_{(#2)} }{            {A} } } }
\DeclareDocumentCommand{\Fg}{ s o }{ \IfBooleanTF{#1}
    { \IfValueTF{#2}{ \bm{\mathcal{F}}_{(#2)} }{ \bm{\mathcal{F}} } }
    { \IfValueTF{#2}{    {\mathcal{F}}_{(#2)} }{    {\mathcal{F}} } } }
\DeclareDocumentCommand{\Ff}{ s o }{ \IfBooleanTF{#1}
    { \IfValueTF{#2}{ \boldsymbol{F}_{(#2)} }{ \boldsymbol{F} } }
    { \IfValueTF{#2}{            {F}_{(#2)} }{            {F} } } }

\DeclareDocumentCommand{\PB}{ O{m} O{q} O{p} m m }{ \frac{ \partial #4 }{\partial {#2}^{#1} } \frac{ \partial #5 }{\partial {#3}_{#1} } - \frac{ \partial #4 }{\partial {#3}_{#1} } \frac{ \partial #5 }{\partial {#2}^{#1} } }

\DeclareDocumentCommand\Te{o o m }{\mathcal{T}{}^{#1}_{#2}(#3)}

\DeclareDocumentCommand{\BM}{ s }{ \IfBooleanTF{#1} {\hat{\bm{M}}}{\bm{M}} }
\DeclareDocumentCommand{\BN}{ s }{ \IfBooleanTF{#1} {\hat{\bm{N}}}{\bm{N}} }
\DeclareDocumentCommand{\BP}{ s }{ \IfBooleanTF{#1} {\hat{\bm{P}}}{\bm{P}} }
\DeclareDocumentCommand{\BQ}{ s }{ \IfBooleanTF{#1} {\hat{\bm{Q}}}{\bm{Q}} }
\DeclareDocumentCommand{\BR}{ s }{ \IfBooleanTF{#1} {\hat{\bm{R}}}{\bm{R}} }
\DeclareDocumentCommand{\BS}{ s }{ \IfBooleanTF{#1} {\hat{\bm{S}}}{\bm{S}} }
\DeclareDocumentCommand{\BU}{ s }{ \IfBooleanTF{#1} {\hat{\bm{U}}}{\bm{U}} }
\DeclareDocumentCommand{\BV}{ s }{ \IfBooleanTF{#1} {\hat{\bm{V}}}{\bm{V}} }

\NewDocumentCommand\MyAc{ m }{#1}
\DeclareDocumentCommand{\vif}{ t. t, t- s s m }{
  \RenewDocumentCommand\MyAc{ m }{##1}
  \IfBooleanT{#1}{\RenewDocumentCommand\MyAc{ m }{ \mathring{##1} } }
  \IfBooleanT{#2}{\RenewDocumentCommand\MyAc{ m }{ \tilde{##1} } }
  \IfBooleanT{#3}{\RenewDocumentCommand\MyAc{ m }{ \bar{##1} } }
  \IfBooleanTF{#4}
  { \IfBooleanTF{#5} { \hat{\MyAc{\boldsymbol{e}}}^{\hat{#6}} }{ \hat{\MyAc{\boldsymbol{e}}}^{{#6}} } }
  { \MyAc{\boldsymbol{e}}^{{#6}} } }
\DeclareDocumentCommand{\vi}{ t. t, t- s s m m}{
  \RenewDocumentCommand\MyAc{ m }{##1}
  \IfBooleanT{#1}{\RenewDocumentCommand\MyAc{ m }{ \mathring{##1} } }
  \IfBooleanT{#2}{\RenewDocumentCommand\MyAc{ m }{ \tilde{##1} } }
  \IfBooleanT{#3}{\RenewDocumentCommand\MyAc{ m }{ \bar{##1} } }
  \IfBooleanTF{#4}
  { \IfBooleanTF{#5} { \hat{\MyAc{e}}^{\hat{#6}}_{\hat{#7}} }{ \hat{\MyAc{e}}^{#6}_{{#7}} } }
  { \MyAc{e}^{{#6}}_{{#7}} } }

\DeclareDocumentCommand{\ct}{ t. t, t- s s m m m }{
  \RenewDocumentCommand\MyAc{ m }{##1}
  \IfBooleanT{#1}{\RenewDocumentCommand\MyAc{ m }{ \mathring{##1} } }
  \IfBooleanT{#2}{\RenewDocumentCommand\MyAc{ m }{ \tilde{##1} } }
  \IfBooleanT{#3}{\RenewDocumentCommand\MyAc{ m }{ \bar{##1} } }
  \IfBooleanTF{#4}
  { \IfBooleanTF{#5} { \hat{\MyAc{\Gamma}}_{{#6}}{}^{\hat{#7}}{}_{\hat{#8}} }{ \hat{\MyAc{\Gamma}}_{{#6}}{}^{{#7}}{}_{{#8}} } }
  { \MyAc{\Gamma}_{{#6}}{}^{{#7}}{}_{{#8}} } }
\DeclareDocumentCommand{\spif}{ t. t, t- s s m m }{
  \RenewDocumentCommand\MyAc{ m }{##1}
  \IfBooleanT{#1}{\RenewDocumentCommand\MyAc{ m }{ \mathring{##1} } }
  \IfBooleanT{#2}{\RenewDocumentCommand\MyAc{ m }{ \tilde{##1} } }
  \IfBooleanT{#3}{\RenewDocumentCommand\MyAc{ m }{ \bar{##1} } }
  \IfBooleanTF{#4}
  { \IfBooleanTF{#5} { \hat{\MyAc{\boldsymbol{\omega}}}^{\hat{#6}}{}_{\hat{#7}} }{ \hat{\MyAc{\boldsymbol{\omega}}}^{{#6}}{}_{{#7}} } }
  { \MyAc{\boldsymbol{\omega}}^{{#6}}{}_{{#7}} } }
\DeclareDocumentCommand{\spi}{ t. t, t- s s m m m }{
  \RenewDocumentCommand\MyAc{ m }{##1}
  \IfBooleanT{#1}{\RenewDocumentCommand\MyAc{ m }{ \mathring{##1} } }
  \IfBooleanT{#2}{\RenewDocumentCommand\MyAc{ m }{ \tilde{##1} } }
  \IfBooleanT{#3}{\RenewDocumentCommand\MyAc{ m }{ \bar{##1} } }
  \IfBooleanTF{#4}
  { \IfBooleanTF{#5} { \hat{\MyAc{{\omega}}}_{\hat{#6}}{}^{\hat{#7}}{}_{\hat{#8}} }{ \hat{\MyAc{{\omega}}}_{{#6}}{}^{{#7}}{}_{{#8}} } }
  { \MyAc{{\omega}}_{{#6}}{}^{{#7}}{}_{{#8}} } }

\DeclareDocumentCommand{\rif}{ t. t, t- s s m m }{
  \RenewDocumentCommand\MyAc{ m }{##1}
  \IfBooleanT{#1}{\RenewDocumentCommand\MyAc{ m }{ \mathring{##1} } }
  \IfBooleanT{#2}{\RenewDocumentCommand\MyAc{ m }{ \tilde{##1} } }
  \IfBooleanT{#3}{\RenewDocumentCommand\MyAc{ m }{ \bar{##1} } }
  \IfBooleanTF{#4}
  { \IfBooleanTF{#5} { \hat{\MyAc{\bm{\mathcal{R}}}}{}^{\hat{#6}}{}_{\hat{#7}} }{ \hat{\MyAc{\bm{\mathcal{R}}}}{}^{{#6}}{}_{{#7}} } }
  { \MyAc{\bm{\mathcal{R}}}{}^{{#6}}{}_{{#7}} } }
\DeclareDocumentCommand{\ri}{ t. t, t- s s m m m }{
  \RenewDocumentCommand\MyAc{ m }{##1}
  \IfBooleanT{#1}{\RenewDocumentCommand\MyAc{ m }{ \mathring{##1} } }
  \IfBooleanT{#2}{\RenewDocumentCommand\MyAc{ m }{ \tilde{##1} } }
  \IfBooleanT{#3}{\RenewDocumentCommand\MyAc{ m }{ \bar{##1} } }
  \IfBooleanTF{#4}
  { \IfBooleanTF{#5} { \hat{\MyAc{\mathcal{R}}}_{{#6}}{}^{\hat{#7}}{}_{\hat{#8}} }{ \hat{\MyAc{\mathcal{R}}}_{{#6}}{}^{{#7}}{}_{{#8}} } }
  { \MyAc{\mathcal{R}}_{{#6}}{}^{{#7}}{}_{{#8}} } }

\DeclareDocumentCommand{\kf}{ t. t, t- s s m m }{
  \RenewDocumentCommand\MyAc{ m }{##1}
  \IfBooleanT{#1}{\RenewDocumentCommand\MyAc{ m }{ \mathring{##1} } }
  \IfBooleanT{#2}{\RenewDocumentCommand\MyAc{ m }{ \tilde{##1} } }
  \IfBooleanT{#3}{\RenewDocumentCommand\MyAc{ m }{ \bar{##1} } }
  \IfBooleanTF{#4}
  { \IfBooleanTF{#5} { \hat{\MyAc{\bm{\mathcal{K}}}}^{\hat{#6}}{}_{\hat{#7}} }{ \hat{\MyAc{\bm{\mathcal{K}}}}^{{#6}}{}_{{#7}} } }
  { \MyAc{\bm{\mathcal{K}}}^{{#6}}{}_{{#7}} } }
\DeclareDocumentCommand{\ko}{ t. t, t- s s m m m }{
  \RenewDocumentCommand\MyAc{ m }{##1}
  \IfBooleanT{#1}{\RenewDocumentCommand\MyAc{ m }{ \mathring{##1} } }
  \IfBooleanT{#2}{\RenewDocumentCommand\MyAc{ m }{ \tilde{##1} } }
  \IfBooleanT{#3}{\RenewDocumentCommand\MyAc{ m }{ \bar{##1} } }
  \IfBooleanTF{#4}
  { \IfBooleanTF{#5} { \hat{\MyAc{\mathcal{K}}}_{{#6}}{}^{\hat{#7}}{}_{\hat{#8}} }{ \hat{\MyAc{\mathcal{K}}}_{{#6}}{}^{{#7}}{}_{{#8}} } }
  { \MyAc{\mathcal{K}}_{{#6}}{}^{{#7}}{}_{{#8}} } }

\DeclareDocumentCommand{\tf}{ t. t, t- s s m }{
  \RenewDocumentCommand\MyAc{ m }{##1}
  \IfBooleanT{#1}{\RenewDocumentCommand\MyAc{ m }{ \mathring{##1} } }
  \IfBooleanT{#2}{\RenewDocumentCommand\MyAc{ m }{ \tilde{##1} } }
  \IfBooleanT{#3}{\RenewDocumentCommand\MyAc{ m }{ \bar{##1} } }
  \IfBooleanTF{#4}
  { \IfBooleanTF{#5} { \hat{\MyAc{\bm{\mathcal{T}}}}^{\hat{#6}} }{ \hat{\MyAc{\bm{\mathcal{T}}}}^{{#6}} } }
  { \MyAc{\bm{\mathcal{T}}}^{{#6}} } }

\newcommand*{\diag}{\operatorname{diag}}

\newcommand{\beq}{\begin{equation}}
\newcommand{\eeq}{\end{equation}}
\newcommand{\ber}{\begin{eqnarray}}
\newcommand{\eer}{\end{eqnarray}}

\newcommand{\dn}[2]{{\mathrm{d}}^{#1}\!{#2}\;}

\newcommand*{\de}[1]{\mathop{\mathrm{d}#1}\nolimits}

\newcommand\UTFSM{Departamento de F\'isica, Universidad T\'{e}cnica Federico Santa Mar\'\i a, \\ Casilla 110-V, Valpara\'iso, Chile}

\newcommand\CCTVal{Centro Cient\'ifico Tecnol\'ogico de Valpara\'iso, \\ Casilla 110-V, Valpara\'\i so, Chile}

\newcommand\PUCV{Instituto de F\'isica, Pontificia Universidad Cat\'olica de Valpara\'iso,\\ Casilla 4059, Valpara\'iso, Chile}

\hypersetup{%
  pdftitle={Kaluza--Klein Cosmology from five-dimensional Lovelock--Cartan Theory},
  pdfauthor={Oscar Castillo-Felisola,}{Cristobal Corral,}{Simon del Pino.},
  pdfkeywords={Torsion,} {Cosmology,}{Kaluza--Klein reduction,}{Generalised Gravity.},
  pdflang={English}
}

\begin{document}

\title{Kaluza-Klein cosmology from five-dimensional Lovelock-Cartan theory}

\author{Oscar \surname{Castillo-Felisola}}
\email{o.castillo.felisola@gmail.com}
\affiliation{\UTFSM.}
\affiliation{\CCTVal.}

\author{Crist\'obal \surname{Corral}}
\email{cristobal.corral@usm.cl}
\affiliation{\UTFSM.}
\affiliation{\CCTVal.}

\author{Sim\'on \surname{del~Pino}}
\email{simon.delpino.m@mail.pucv.cl}
\affiliation{\PUCV.}

\author{Francisca \surname{Ram\'irez}.}
\email{francisca.ramirez@alumnos.usm.cl}
\affiliation{\UTFSM.}

\begin{abstract}
  We study the Kaluza-Klein dimensional reduction of the Lovelock-Cartan theory in five-dimensional spacetime, with a compact dimension of $S^1$ topology. We find cosmological solutions of the Friedmann-Robertson-Walker class in the reduced spacetime. The torsion and the fields arising from the dimensional reduction induce a nonvanishing energy-momentum tensor in four dimensions. We find solutions describing expanding, contracting, and bouncing universes. The model shows a dynamical compactification of the extra dimension in some regions of the parameter space. 
\end{abstract}


\maketitle

\section{Introduction}

The experimental status of general relativity (GR), regarding the solar system tests~\cite{Will:2014kxa} and the detection of signals consistent with the merge of two black holes by the LIGO Collaboration~\cite{Abbott:2016blz,*Abbott:2016nmj}, has settled it as the most successful theory of gravity. However, the difficulties in finding explanations for the so-called dark sector of the Universe have driven the community to think that GR is not the ultimate gravitational theory. 

The dark sector of the Universe is composed by two kinds of degrees of freedom. On the one hand, the gravitational lensing produced by the local distribution of energy suggests the presence of an exotic form of matter unseen by the current light-based telescopes~\cite{Wittman:2000tc}. It is thus named dark matter and it would interact mostly (if not only) through gravity. Such an abundance at the galactic scale is compatible with the velocity profile of stars at its outer regions~\cite{Sofue:2000jx}. At the cosmic scale, it plays a key role in the origin and evolution of structures (see, for instance Ref.~\cite{DelPopolo:2008mr}). On the other hand, the experimental data obtained from type Ia supernovae observations, indicates that our Universe is passing through a phase of accelerated expansion~\cite{Riess:1998cb}. This behavior suggests the existence of an exotic form of energy, called dark energy, which constitutes roughly $70\%$ of the current content of our Universe.  

The shortcomings of GR on describing these phenomena are the main motivation to look for new gravitational degrees of freedom. Among the possible extensions, higher-dimensional models could shed some light on the nature of these new degrees of freedom. For instance, as was shown in the early works of Kaluza and Klein, the existence of an extra dimension within the GR framework would give rise to a unified picture of gravity and electromagnetism, along with a spectrum of new heavy particles~\cite{Kaluza:1921tu,*Klein:1926tv}. This idea opened the possibility of a novel geometrical understanding of interactions, where the gauge group arises as a consequence of the topology of the spatial compact manifold in a higher-dimensional spacetime. The idea of higher dimensions comes naturally in diverse physical models, for example: supersymmetry and supergravity~\cite{supergravity}, string theory~\cite{Green:1987sp,*Green:1987mn}, novel proposals by Arkani-Hammed \emph{et al.}~\cite{ArkaniHamed:1998rs,*Antoniadis:1998ig}, and models from Randall and Sundrum~\cite{Randall:1999ee,*Randall:1999vf}, as attempts to solve the hierarchy problem.

In four dimensions, the Einstein-Hilbert action with a cosmological constant is the most general theory which leads to second-order field equations for the metric. In higher dimensions, however, particular combinations of higher-order terms in the curvature can be added to the gravitational action, whose variation with respect to the metric also yields to second-order field equations. The most general theory in arbitrary dimensions, which preserves this feature of the four-dimensional Einstein-Hilbert action, is called the Lanczos-Lovelock action~\cite{Lanczos:1938sf,*Lovelock:1971yv}. Such a theory has no ghosts~\cite{Zumino:1985dp} and has the same degrees of freedom as the Einstein-Hilbert Lagrangian in arbitrary dimensions~\cite{Henneaux:1990au}. It is worth mentioning that in the Palatini approach, where the metric and the connection are considered as independent fields, there are families of Lagrangians which yield to second-order field equations, and do not possess ghosts~\cite{Olmo:2011uz}.

The simplest possible extra term in the Lanczos--Lovelock action  is a quadratic construction of curvatures, called the Gauss-Bonnet term, which reads
\begin{equation}\label{GB}
  \mathcal{L}_{\mathrm{GB}} = \dn{N}{x}\sqrt{-g}\left(\tilde{R}^2-4\tilde{R}_{\mu\nu}\tilde{R}^{\mu\nu}+\tilde{R}_{\alpha\beta\mu\nu}
  \tilde{R}^{\alpha\beta\mu\nu}\right),
\end{equation}
where $\tilde{R}_{\alpha\beta\mu\nu}$ is the Riemannian curvature of a manifold with metric $g_{\mu\nu}$ and $g$ its determinant. $\tilde{R}_{\mu\nu}$ and $\tilde{R}$ are the Ricci tensor and Ricci scalar, respectively. In four dimensions, the Gauss-Bonnet term adds no dynamics to the metric, since it represents a topological invariant proportional to the Euler characteristic class, which can be written locally as a boundary term. Nevertheless, it has been shown that this term can be relevant for conserved charges considerations in spacetimes with a local AdS asymptotic~\cite{Aros:1999id}. Moreover, as it was reported in Ref.~\cite{Miskovic:2009bm}, the inclusion of topological invariants of the Euler class is equivalent to the program of holographic renormalization in the context of AdS/CFT. In dimensions higher than four it contributes to the field equations, and it was also identified as the low-energy correction for a spin-two field in string theory~\cite{Zwiebach:1985uq}.

The Riemannian five-dimensional Lanczos-Lovelock theory has been widely studied in the literature. For instance, exact wormhole solutions which violate no energy conditions have been found in vacuum~\cite{Dotti:2006cp,*Dotti:2007az}, and coupled with matter fields satisfying the weak energy conditions~\cite{Mehdizadeh:2015jra}. In higher-order Lanczos-Lovelock models, this class of exact solutions have been reported in~\cite{Mehdizadeh:2015dta} and in the compactified theory with torsion~\cite{Canfora:2008ka}. Additionally, the compactification of higher Lanczos-Lovelock terms has been considered in Ref.~\cite{MuellerHoissen:1985mm,*MuellerHoissen:1989yv} and their cosmology in
Refs.~\cite{MuellerHoissen:1985ij,Deruelle:1986iv,Deruelle:1989fj}. It is worth mentioning that classically, in GR, wormholes must be supported by a kind of energy compatible with the cosmological hypothesis. Their existence in these extended models implies the presence of degrees of freedom that could provide an explanation for the exotic matter/energy abundance in the Universe.

On the other hand, it is well known that GR assume the torsion-free condition {\it a priori}. However, as Cartan first considered, it is possible to take metricity and parallelism as truly independent concepts~\cite{Cartan1922,*Cartan1924,*Cartan1925}. Such kinds of geometry are known as Riemann--Cartan geometries and offer the natural framework for Poincar\'e gauge theories, where the torsion appears as the field strength of translations, sourced by the spin current~\cite{Kibble:1961ba,Sciama:1962,Hehl:1976kj}.
Moreover, the vacuum predictions of its simplest formulation---the Einstein-Cartan theory---holds the experimental tests of GR. In the same spirit of the Gauss--Bonnet extension of GR, the quadratic corrections in curvature and torsion to Einstein--Cartan have been considered, see, for example Refs.~\cite{Baekler:2011jt,Fabbri:2012qr}. The cosmological consequences of non-Riemannian geometries has been widely studied in the literature (for a review, see Ref.~\cite{Puetzfeld:2004yg}).

In the framework of the Kaluza-Klein theories in higher-dimensional Riemann-Cartan geometries, phenomenology of the extradimensional torsion was found in Ref.~\cite{Kalinowski:1980da}, and metric-dependent torsion in extra dimensions was studied in~\cite{Shankar:2012vd}, along with its consequences in cosmology~\cite{Chen:2009ep}. The compatification of higher-dimensional Brans-Dicke models with torsion was considered in Ref.~\cite{German:1993bq}. Torsion-free black-hole solutions were found in~\cite{Aros:2007nn}, for first-order compactified gravity.

The extension of the Lanczos-Lovelock theories with nonvanishing torsion is known as Lovelock-Cartan theory~\cite{Mardones:1990qc}. In that framework, we present a new class of cosmological solutions in five dimensions, where the compact dimension is $S^1$. The theory admits a nonvanishing torsion in vacuum due to the presence of the Gauss-Bonnet term, in contrast to the five-dimensional Einstein-Cartan theory. The new degrees of freedom coming from the higher-dimensional geometry generate an induced energy-momentum form in the reduced theory. In some cases, the solutions avoid the
appearance of initial singularities and drive the accelerated expansion of the Universe, while the radius of the compact manifold goes to zero.

This work is organized as follows. In Sec.~\ref{KK} we review the Kaluza-Klein (KK) geometry in the first-order formalism and fix our notation and conventions. In Sec.~\ref{5EGB}, we study the dynamics of the general Lovelock-Cartan action in a five-dimensional spacetime and its dimensional reduction. In Sec.~\ref{cosmos}, we look for cosmological solutions of the Friedmann-Robertson-Walker class with nontrivial torsion. Conclusions and remarks are given in Sec.~\ref{conclusions}. We also incorporate a number of appendices to make this work a self-contained article.

\section{Kaluza-Klein Geometry in Riemann-Cartan Spacetimes\label{KK}}

In the following we will consider $M_N$ to be a $N$-dimensional differential manifold. Every quantity defined on $M_N$ will be denoted by hats $\hat{x}$. Capital greek characters (coordinate indices) and capital latin characters (Lorentz indices) run over the $N$ dimensions, i.e., \mbox{$A = 0,...,N-2,N$}, while the lowercase ones run in the $(N-1)$-dimensional reduced manifold, i.e., \mbox{$a = 0,...,N-2$}.

The differential structure of $M_N$ is determined by two fields. The spin connection 1-form, $\hat{\omega}^{AB} = \hat{\omega}^{AB}{}_{\Gamma}\,\text{d}\hat{x}^\Gamma$, describes the affine structure of the $N$-dimensional manifold and the vielbein 1-form, $\hat{e}^A=\hat{e}^{A}{}_{\Gamma}\,\text{d}\hat{x}^\Gamma$, defines the metric structure of the same manifold through the relation $\hat{g}_{\Gamma\Delta} = \hat{e}^{A}{}_{\Gamma}\hat{e}^{B}{}_{\Delta}\hat{\eta}_{AB}$, where $\hat{g}_{\Gamma\Delta}$ is the spacetime metric, while $\hat{\eta}_{AB} = \diag(-,+,...,+)$ is the Minkowski metric. The vielbein maps coordinate into Lorentz indices.
In our convention, the Levi-Civita symbol is such that $\hat{\epsilon}_{01...(N-2)N} = 1$ and the reduced symbol is defined by fixing the last index, i.~e.
\[\hat{\epsilon}_{a_1 \cdots a_{N-1} N} \equiv \epsilon_{a_1 \cdots a_{N-1}}.\]
Additionally, all the Riemannian fields (torsion free) will be made explicit by a tilde, as it was done in Eq.~\eqref{GB} for the curvature.

Because of the topology of the manifold, we can expand the dependence of the fields on the extra coordinate, $z$, in a Fourier series as  
\begin{equation}
  \label{Fourier}
  \hat{\varrho}(\hat{x}^\Gamma)=\sum_n\varrho_{(n)}(x)e^{i n z},
\end{equation}
where $x$ denotes the coordinates of the $(N-1)$-manifold, collectively. Henceforth, we will focus on the $n=0$ mode of the expansion, also referred to as the low-energy sector.

\subsection{Metric and affine structure}

The KK ansatz for the metric lies in the premise that the compact dimension of $M_N$ is orthogonal to the rest of the manifold at each point. This leads to the following metric structure:
\begin{equation}
  \hat{g}_{\Gamma\Delta} =
  \begin{pmatrix}
    g_{\gamma\delta} +\frac{\hat{g}_{\gamma z}\hat{g}_{\delta z}}{\hat{g}_{zz}}&\hat{g}_{\gamma z}\\
    \hat{g}_{z\delta} & \hat{g}_{zz}
  \end{pmatrix}
  =
  \begin{pmatrix}
    g_{\gamma\delta} + \phi A_\gamma A_\delta&\phi A_\gamma\\
    \phi A_{\delta} & \phi
  \end{pmatrix}
\end{equation}
and its inverse,
\begin{equation}
  \hat{g}^{\Gamma\Delta}=
  \begin{pmatrix}
    g^{\gamma\delta}&-A^\gamma\\
    -A^{\delta} & \phi^{-1}+A^2
  \end{pmatrix},
\end{equation}
defined such that $\hat{g}_{\Gamma\Delta}\hat{g}^{\Delta\Lambda}=\delta^\Lambda_\Gamma$. This metric introduces a scalar field $\phi$ and a vector field $A_\mu$ as new gravitational degrees of freedom. The vielbein that holds this structure for the metric has the following form:
\begin{equation}
  \label{Dvielbein}
  \hat{e}^A{}_{\Gamma} =
  \begin{pmatrix}
    \hat{e}^a{}_{\gamma}& 0\\
    \hat{e}^N{}_{\gamma} & \hat{e}^N{}_{z}
  \end{pmatrix}
  =
  \begin{pmatrix}
    e^a{}_{\gamma}& 0\\
    \sqrt{\phi}A_\gamma & \sqrt{\phi}
  \end{pmatrix},
\end{equation}
and its inverse, $\hat{E}_A{}^{\Gamma}$, defined such that $\hat{E}_A{}^{\Gamma}\hat{e}^A{}_{\Delta}=\delta^\Gamma_{\Delta}$ and $\hat{E}_A{}^{\Gamma}\hat{e}^B{}_{\Gamma}=\delta^B_A$, reads
\begin{equation}
  \label{Dinversevielbein}
  \hat{E}_A{}^{\Gamma} =
  \begin{pmatrix}
    \hat{E}_a{}^{\gamma}& 0\\
    \hat{E}_a{}^{z} & \hat{E}_N{}^{z}
  \end{pmatrix}
  =
  \begin{pmatrix}
    E_a{}^{\gamma}& 0\\
    -A_a & \sqrt{\phi}^{-1}
  \end{pmatrix},
\end{equation}
where $A_a = E_a{}^\mu A_\mu$. This shows that the compact manifold $S^1$ has its own tangent space, $T_pS^1$, at each point $p\in M_N$, and is independent of $T_pM_{N-1}$ in the sense that they do not mix. 

The vielbeins are defined modulo Lorentz transformations. Any transformed basis $\hat{e}^{\prime A} = \hat{\Lambda}^A{}_{B}\,\hat{e}^B$,
where $\hat{\Lambda}$ is a Lorentz matrix, is as suitable as $\hat{e}^A$, and therefore shares the structure of Eq.~\eqref{Dvielbein}. The Lorentz transformations are then constrained by $\hat{\Lambda}^a{}_{N}=0$.

A Riemannian connection compatible with the \mbox{$N$-dimensional} vielbein~\eqref{Dvielbein}, is built under the premise that $\mbox{d}\hat{e}^A + \hat{\tilde{\omega}}^{A}{}_B \wedge \hat{e}^B = 0$. Thus, we find
\begin{equation}
  \label{DRiemannconnection}
  \begin{split}
    \hat{\tilde{\omega}}^{ab}&=\tilde{\omega}^{ab}-\frac{1}{2}\sqrt{\phi}F^{ab}\hat{e}^N,\\
    \hat{\tilde{\omega}}^{Na}&=\frac{1}{2}\sqrt{\phi}F^a_{\ \ l}e^l+\frac{1}{2}\partial^a\ln\phi\hat{e}^N,
  \end{split}
\end{equation}
where $\tilde{\omega}^{ab}$ is the Riemannian spin connection of the reduced manifold and $F_{ab}$ are the components of the field strength of $A = A_m\, e^m$, defined by
\begin{equation}
  F=\text{d}A=\frac{1}{2}F_{ab}\, e^a\wedge e^b.
\end{equation}
The construction of the five-dimensional Einstein-Hilbert action by means of the spin connection~\eqref{DRiemannconnection} leads to the original KK theory.

In a Riemann-Cartan geometry, $M_N$ is not entirely described by $\hat{\tilde{\omega}}^{AB}$, but by a more general spin connection independent of the metric degrees of freedom (see Appendix~\ref{Riemann-Cartan} for details). We can assume a general connection 1-form of the same type (regarding its $M_{N-1}\times S^1$ decomposition) as in Eq.~\eqref{DRiemannconnection}. Thus, the most general spin connection on $M_N$ compatible with the KK decomposition is given by 
\begin{equation}\label{Nconnection}
  \hat{\omega}^{AB} \equiv
  \begin{pmatrix}
    \omega^{ab}+\alpha^{ab}\hat{e}^N & \beta^a+\gamma^a\hat{e}^N\\
    -\beta^b-\gamma^b\hat{e}^N & 0
  \end{pmatrix}.
\end{equation}

The decomposition in Eq.~\eqref{Nconnection} adds new metric-independent fields. The 0-form $\alpha^{ab}$ is an antisymmetric tensor of spin-1. The 1-form $\beta^a = \beta^a{}_\mu \de{x}^\mu$ generically adds a spin-$2$ field, a spin-$1$ field and a spin-$0$ field. The last piece, $\gamma^a$, is a vector $0$-form of spin-1.

\subsection{Curvature and torsion}

The $N$-dimensional Lorentz curvature and torsion are given by the Cartan structure equations
\begin{align}
  \label{curvadef}
  \hat{R}^{AB} &= \mbox{d}\hat{\omega}^{AB}+\hat{\omega}^A_{\ \ C}\wedge\hat{\omega}^{CB} = \frac{1}{2} \hat{R}^{AB}{}_{CD} \; \hat{e}^C \wedge \hat{e}^D,\\
  \label{tordef}
  \hat{T}^A &= \mbox{d}\hat{e}^A+\hat{\omega}^A_{\ \ B}\wedge\hat{e}^B = \frac{1}{2} \hat{T}^{A}{}_{BC} \; \hat{e}^B \wedge \hat{e}^C. 
\end{align}
Using the definition of curvature in Eq.~\eqref{curvadef},  with the KK ansatz for the spin connection~\eqref{Nconnection}, we find
\begin{align}
  \label{R1}
  \hat{R}^{ab}&=R^{ab}+\sqrt{\phi}\alpha^{ab}F-\beta^a\wedge\beta^b\notag\\
  & \quad +\left(\mbox{D}\alpha^{ab}+\frac{1}{2}\alpha^{ab}\mbox{d}\ln\phi-2\beta^{[a}\gamma^{b]}\right)\wedge\hat{e}^N,\\
  \label{R2}
  \hat{R}^{Na}&=-\left(\mbox{D}\beta^a+\sqrt{\phi}\gamma^a F\right)\notag\\
  & \quad +\left(\alpha^a_{\ b}\beta^b-\mbox{D}\gamma^a-\frac{1}{2}\gamma^a\mbox{d}\ln\phi\right)\wedge\hat{e}^N,
\end{align}
where $R=\text{d}\omega+\omega\wedge\omega$ is the curvature of the reduced spacetime $M_{N-1}$. Similarly from the definition of torsion in Eq.~\eqref{tordef}, we find its distinctive parts to be\footnote{Henceforth we will refer to the distinctive parts concerning the $M_{N-1}\times S^1$ decomposition.}
\begin{align}\label{T1}
  \hat{T}^a &= T^a+\left(\beta^a-\alpha^a_{\ \ b}e^b\right)\wedge\hat{e}^N,\\
  \label{T2}
  \hat{T}^N &= \sqrt{\phi}F-\beta_b\wedge e^b+\left(\frac{1}{2}\mbox{d}\ln\phi+\gamma_be^b\right)\wedge\hat{e}^N,
\end{align}
where $T=\text{d}e+\omega\wedge e$ is the torsion $2$-form of $M_{N-1}$.

\subsection{Bianchi identities\label{sec:bianchi}}

Considering the Bianchi identities for the KK structure described above, we find relevant information about the new fields. Taking the exterior covariant derivative over the $N$-dimensional curvature and torsion, the Bianchi identities are
\begin{equation}
  \hat{\text{D}}\hat{R}^{AB} = 0 \quad \text{and} \quad \hat{\text{D}}\hat{T}^A = \hat{R}^A_{\ B}\wedge\hat{e}^B.
\end{equation}
A careful decomposition of the first Bianchi identity (covariant derivative of the curvature above) into its distinctive parts gives the Bianchi identity for the curvature of the reduced spacetime, together with the second derivative rules
\begin{equation}
  \begin{aligned}
    \text{D}R^{ab} &=0, & \text{D}\mbox{D}\alpha^{ab} &=R^a_{\ l}\alpha^{lb}+R^b_{\ l}\alpha^{al},\\
    \text{D}\text{D}\beta^a &= R^a_{\ b}\beta^{b}, & \text{D}\text{D}\gamma^a &= R^a_{\ b}\gamma^{b}.
  \end{aligned}
\end{equation}
This is taken as a proof of the tensorial nature of these new fields under the four-dimensional Lorentz transformations.\footnote{The tensorial nature of these fields can be also derived from the decomposition of the transformation rule of $\hat{\omega}^{AB}$ under the Lorentz group.} The second Bianchi identity gives its equivalent for $M_{N-1}$; this is
\begin{equation*}
  \mbox{D}T^a=R^a_{\ b}\wedge e^b.
\end{equation*}

\section{Five-Dimensional Lovelock-Cartan Reduction\label{5EGB}}

The most general theory requiring the Lagrangian to be (i)~an invariant $N$-form under local Lorentz transformations, (ii)~a local polynomial of the vielbein, the Lorentz connection, and their exterior derivatives, and (iii)~constructed without the Hodge dual,\footnote{The Hodge dual maps $p$-forms into $(N-p)$-forms through $\star\left(\hat{e}^{A_1}\wedge ... \wedge\hat{e}^{A_p}\right) = \frac{1}{(N-p)!}\hat{\epsilon}^{A_1\ldots A_p}{}_{A_{p+1}...A_N}\,\hat{e}^{A_{p+1}}\wedge ... \wedge\hat{e}^{A_N}$.} is the Lovelock-Cartan theory of gravity~\cite{Mardones:1990qc}. This is the natural generalization of Lanczos-Lovelock action when torsional degrees of freedom are present. Its simplest realization is the Einstein-Cartan model and their dimensional reduction is indistinguishable from the Riemannian case (see, for instance,~\cite{German:1993bq,Aros:2007nn}). 

In five dimensions, the Lovelock-Cartan theory is given by the following action principle 
\begin{widetext}
  \begin{equation}
    \label{action5EGB}
    I = \int_{M_5} \hat{\epsilon}_{ABCDE} \Big(\frac{\alpha_0}{5}\hat{e}^A\wedge\hat{e}^B\wedge\hat{e}^C\wedge
    \hat{e}^D\wedge\hat{e}^E
    +\frac{\alpha_1}{3}\hat{R}^{AB}\wedge\hat{e}^C\wedge\hat{e}^D\wedge\hat{e}^E
    +\alpha_2\hat{R}^{AB}\wedge\hat{R}^{CD}
    \wedge\hat{e}^E\Big),
  \end{equation}
  where $\alpha_0$, $\alpha_1$, and $\alpha_2$ are dimensionful coupling constants. Its variation with respect to the five-dimensional vielbein and spin connection give the equations
  \begin{gather}
    \label{delta e}
    \hat{\epsilon}_{ABCDE}\Big(\alpha_0\hat{e}^B\wedge\hat{e}^C\wedge\hat{e}^D\wedge\hat{e}^E
    + \alpha_1\hat{R}^{BC}\wedge\hat{e}^D\wedge\hat{e}^E
    + \alpha_2\hat{R}^{BC}\wedge\hat{R}^{DE}\Big)=0,
    \\
    \label{delta w}
    \hat{\epsilon}_{ABCDE}\Big(\alpha_1\hat{e}^C\wedge\hat{e}^D+
    2\alpha_2\hat{R}^{CD}\Big)\wedge\hat{T}^E=0.
  \end{gather}
\end{widetext}

The last term in Eq.~\eqref{action5EGB} is the five-dimensional Gauss-Bonnet term for a Lorentz curvature written in exterior forms, which is analogous to Eq.~\eqref{GB} for the Riemannian case. Exact solutions with torsion were reported in~\cite{Canfora:2007ux}.

The Lovelock-Cartan theory in five dimensions allows a unique torsional extension that is not in the Lovelock series~\cite{Mardones:1990qc},
\begin{align}
  \label{boundary}
  \mathcal{L}_{T} \propto \hat{T}_A\wedge \hat{R}^A_{\ B}\wedge\hat{e}^B.
\end{align}
However, it can be written as a boundary term, adding no dynamics to the field equations.

In this work, we will focus on the region in the parameter space where
\begin{equation}
  \label{delta}
  \Delta\equiv\alpha_1^2-4\alpha_0\alpha_2 > 0.
\end{equation}
This condition place us outside the Chern-Simons point ($\Delta = 0$). In that particular case, the field equations~\eqref{delta e} and~\eqref{delta w} are invariant under a larger gauge group ($AdS_5$), while the action~\eqref{action5EGB} becomes the Chern-Simons form for that group~\cite{Zanelli:2016cs,*Troncoso:1999pk}. Exact solutions with torsion in Chern-Simons gravity are given in~\cite{Aros:2007nn,Banados:2003cz}. Holographic properties in the Riemannian case were studied in Ref.~\cite{Banados:2005rz} and also, when torsion is considered, in Ref.~\cite{Banados:2006fe}. Additionally, the case $\Delta < 0$ was considered in Ref.~\cite{Canfora:2013xsa}, and its cosmology in Ref.~\cite{Canfora:2014iga}.

In terms of the KK ansatz presented in the previous section, the field equations can be decomposed into its distinctive parts. Then Eq.~\eqref{delta e} leads to 
\setlength\multlinegap{0pt}
\begin{gather}
  \begin{multlined}
    \label{equation}
    \epsilon_{abcd}\Big[\alpha_0 e^b\wedge e^c\wedge e^d + \frac{1}{2}\alpha_1\left(M^{bc}-L^b\wedge e^c\right)\wedge e^d\\
      -\alpha_2\left(L^b\wedge M^{cd} + K^b\wedge N^{cd}\right)\Big]=0,
  \end{multlined}
  \\
  \epsilon_{abcd}K^b\wedge\left(\alpha_1e^c\wedge e^d + 2\alpha_2M^{cd}\right)=0,
  \\
  \begin{multlined}
    \epsilon_{abcd}\big(\alpha_0 e^a\wedge e^b\wedge e^c\wedge e^d \\
    +\alpha_1M^{ab}\wedge e^c\wedge e^d+\alpha_2M^{ab}\wedge M^{cd}\big)=0,
  \end{multlined}
  \\
  \epsilon_{abcd}\left(\alpha_1N^{ab}\wedge e^c\wedge e^d+2\alpha_2N^{ab}M^{cd}\right)=0,
\end{gather}
while Eq.~\eqref{delta w} gives
\setlength\multlinegap{0pt}
\begin{gather}
  \begin{multlined}
    \epsilon_{abcd}\Bigl[\alpha_1\left(e^c\wedge e^d\wedge Z-2e^c\wedge T^d\right)
      +2\alpha_2\big(N^{cd}\wedge W\\
      +M^{cd}\wedge Z+2L^c\wedge T^d
      +2K^c\wedge V^d\big)\Bigr]=0,
  \end{multlined}
  \\
  \begin{multlined}
    \epsilon_{abcd}\Bigl[\alpha_1e^c\wedge e^ d\wedge W \\
      + 2\alpha_2\left(M^{cd}\wedge W+2K^c\wedge T^d\right)\Bigr]=0,
  \end{multlined}
  \\
  \epsilon_{abcd}\left(\alpha_1e^b\wedge e^c+2\alpha_2M^{bc}\right)\wedge T^d=0,
  \\
  \begin{multlined}
    \epsilon_{abcd}\Bigl[\alpha_1e^b\wedge e^c\wedge V^d \\
      + 2\alpha_2\big(M^{bc}\wedge V^d+N^{bc}\wedge T^d\big)\Bigr]=0.
  \end{multlined}
\end{gather}
The fields $M^{ab}$, $N^{ab}$, $L^a$, $K^a$, $W$, $V^a$, and $Z$ are defined from Eqs.~\eqref{R1}--\eqref{T2}, such that
\begin{equation}
  \begin{aligned}
    \hat{R}^{ab}&=M^{ab}+N^{ab}\wedge\hat{e}^5, & \hat{R}^{5a}&=K^a+L^a\wedge\hat{e}^5,\\
    \hat{T}^a&=T^a+V^a\wedge\hat{e}^5, & \hat{T}^5&=W+Z\wedge\hat{e}^5.
  \end{aligned}
\end{equation}

\section{Dimensionally Reduced Lovelock-Cartan Cosmology\label{cosmos}}

\subsection{Cosmological ansatz}

In order to look for cosmological solutions, we demand the symmetries assumed by the cosmological principle, i.e., isotropy and homogeneity of the involved fields. This can be achieved by imposing that the Lie derivative of each field along the Killing vectors that generate the symmetries vanishes. Appendix~\ref{homotropic} is devoted to the details of how to find the general ansatz.
The metric for this case is given by a Friedmann-Robertson-Walker ansatz whose line element reads
\begin{align*}
\text{d}s^2=-\text{d}t^2+a^2(t)\left(\frac{\text{d}r^2}{1-kr^2}+r^2\text{d}\theta^2+r^2\sin^2\varphi\text{d}\varphi^2\right),
\end{align*}
where $a(t)$ is the scale factor and $k=+1,0,-1$ determines the spatial section of the four-manifold to be closed, flat or open, respectively. The four-dimensional vielbein compatible with this metric reads
\begin{equation}
  \begin{aligned}
    \label{vielbein cosmo}
    e^0&=\mbox{d}t, & e^1&=\frac{a(t)}{\sqrt{1-kr^2}} \, \mbox{d}r,\\
    e^2&=a(t)r \, \mbox{d}\theta, & e^3&=a(t) r \sin\varphi \,\mbox{d}\varphi.
  \end{aligned}
\end{equation}
The scalar field is a time-dependent function, 
\begin{equation}
  \phi=\phi(t),
\end{equation}
which is interpreted as the scale factor of the compact extra dimension.

On the other hand, the spin connection adds two functions that we called $\omega(t)$ and $f(t)$,
\begin{align}
  \omega^{0i}&=\omega(t) \, e^i,\\
  \omega^{12}&=-\frac{\sqrt{1-kr^2}}{a(t)r} \, e^2 - f(t) \, e^3,\\
  \omega^{13}&=-\frac{\sqrt{1-kr^2}}{a(t)r} \, e^3 + f(t) \, e^2,\\
  \omega^{23}&=-\frac{\cot\theta}{a(t)r} \, e^3 - f(t) \, e^1.
\end{align}

The nonvanishing components of the spin connection induced by the compact manifold are
\begin{gather}
  \beta^0 = -b(t) \, e^0,\quad \beta^i = \beta(t) \, e^i,\\
  \label{gamma cosmo}
  \gamma^0 =-\gamma(t).
\end{gather}

For this ansatz, the $1$-form $A = A(t)\text{d}t$ has vanishing field strength $F$. Thus, without loss of generality, it can be fixed to zero by means of an $U(1)$ gauge transformation or, equivalently, a diffeomorphism transformation along the $z$ direction.

The equations of motion for the cosmological ansatz give a system of differential equations for the time-dependent functions defined above. Curvature and torsion for this ansatz can be seen in Appendix~\ref{homotropic}. We find that the only non-Riemannian branch demands  $\beta^a=0$. Thus $b(t)=\beta(t)=0$. Otherwise, the system is consistent only at the Chern-Simons point, where it degenerates. 

After a simple algebra, the system becomes
\setlength\multlinegap{0pt}
\begin{gather}
  \label{eqn1}
  \begin{multlined}
    h\left[\alpha_1+2\alpha_2\left(\omega^2+\frac{k}{a^2}-f^2\right)\right] \\
    +4\alpha_2f\left(\dot{f}+Hf\right)=0,
  \end{multlined}
  \\
  \begin{multlined}
    2\alpha_0+\alpha_1\left(\dot{\omega}+2\omega^2+\frac{k}{a^2}-f^2\right)\\
    +2\alpha_2\left(\dot{\omega}+\omega^2\right)\left(\omega^2+\frac{k}{a^2}-f^2\right)=0,
  \end{multlined}
  \\
  2\alpha_2\left(\tfrac{\dot{\phi}}{2\phi}+\gamma\right)f\left(\dot{f}+Hf\right)
  +h^2\left(\alpha_1-2\alpha_2\omega\gamma\right)=0,
  \\
  \omega\left(\frac{\dot{\phi}}{2\phi}+\gamma\right)+\dot{\gamma}+\frac{\dot{\phi}}{2\phi}\gamma-\frac{\alpha_1}{2\alpha_2}=0,
  \\
  \left(\omega^2+\tfrac{k}{a^2}-f^2\right)\left(\alpha_1-2\alpha_2\omega\gamma\right)-\alpha_1\omega\gamma+2\alpha_0=0,
  \\
  \label{eqn2}
  \left(\dot{\omega}+\omega^2\right)\left(\alpha_1-2\alpha_2\omega\gamma\right)-\alpha_1
  \omega\gamma+3\alpha_0 - \frac{\alpha^2_1}{4\alpha_2}=0,
\end{gather}
where dot stands for the time derivative. For the sake of simplicity, we have defined $h(t)=\omega(t)-H(t)$, where $H=\dot{a}/a$ is the Hubble function.

\subsection{Solutions\label{sec:sol}}

The system develops two non-Riemannian branches, one for each value of the parameter
\begin{equation}
  \label{u}
  u_\pm=\frac{2\alpha_1\pm\sqrt{6\Delta}}{4\alpha_2},
\end{equation}
where $\Delta$ was given in Eq.~\eqref{delta}.

Equations \eqref{eqn1}--\eqref{eqn2} are reduced to the Riemannian system when \mbox{$f=h=0$} and \mbox{$\gamma=-\dot{\phi}/2\phi$}. The details of such a model can be seen in Refs.~\cite{Deruelle:1986iv,Deruelle:2003ck,Henriques:1986jw,*Ishihara:1986if,Kleidis:1997mu,Canfora:2016umq}.

Because of Eq.~\eqref{u}, the solutions will be valid in the region of the parameter space where $\Delta>0$. The function $\omega(t)$ satisfies the equation
\begin{equation}
  \dot{\omega}+\omega^2+u_{\pm} = 0,
\end{equation}
which, for the three significantly different values of $u_\pm$, has the following solutions
\begin{equation}
  \omega(t) =
  \begin{cases}
    -\sqrt{u_\pm}\tan\left[\sqrt{u_\pm}\left(t-t_0\right)\right], & u_\pm > 0 \\
    \left(t-t_0\right)^{-1}, & u_\pm = 0 \\
    \sqrt{-u_\pm}\tanh\left[\sqrt{-u_\pm}\left(t-t_0\right)\right], & u_\pm < 0
  \end{cases}
\end{equation}
where $t_0$ is an integration constant to be fixed. We express the time dependence of the remaining fields in terms of $\omega$ and list them explicitly in Appendix~\ref{solutions t}. The solutions read
\begin{align}
  \gamma(\omega)&=\frac{u_\pm}{\omega},\\
  \phi(\omega)&=\frac{\phi_0\,\omega^2}{\left|\omega^2+u_\pm\right|}\exp\left[\frac{\mp\sqrt{6\Delta}}{4\alpha_2\left(\omega^2+u_\pm\right)}\right],\\
  a(\omega)&=\frac{a_0}{\sqrt{\left|\omega^2+u_\pm\right|}}\exp\left[\frac{\pm\sqrt{6\Delta}} {24\alpha_2\left(\omega^2+u_\pm\right)}\right],\\
  \label{f2sol}
  f^2(\omega)&=\left|\omega^2+u_\pm\right|\frac{k}{a_0^2}\exp\left[\frac{\mp\sqrt{6\Delta}}{12\alpha_2\left(\omega^2+u_\pm\right)}\right]\notag\\
  & \quad +\omega^2  + \frac{3\alpha_1\pm\sqrt{6\Delta}}{6\alpha_2},
\end{align}
where $a_0$ and $\phi_0$ are integration constants. 

Herein, we will consider $\alpha_1$ to be positive,\footnote{In all the plots we normalize $\alpha_1 = 1$.} because this parameter admits the interpretation of the Newton's gravitational constant. It is worth mentioning the existence of models of gravity that are free of the linear curvature term. These models are referred to as pure Lovelock gravities~\cite{Cai:2006pq} and consider the cosmological term and the polynomial of highest order in the curvature. Even though the analysis that follows will consider a positive nonzero value for $\alpha_1$, the solutions listed above are also valid in the regime where $\alpha_1=0$, which constitutes the non-Riemannian extension of pure Lovelock theories.

\subsubsection{Bouncing solutions}

\begin{figure}[H]
  \includegraphics[width=\linewidth]{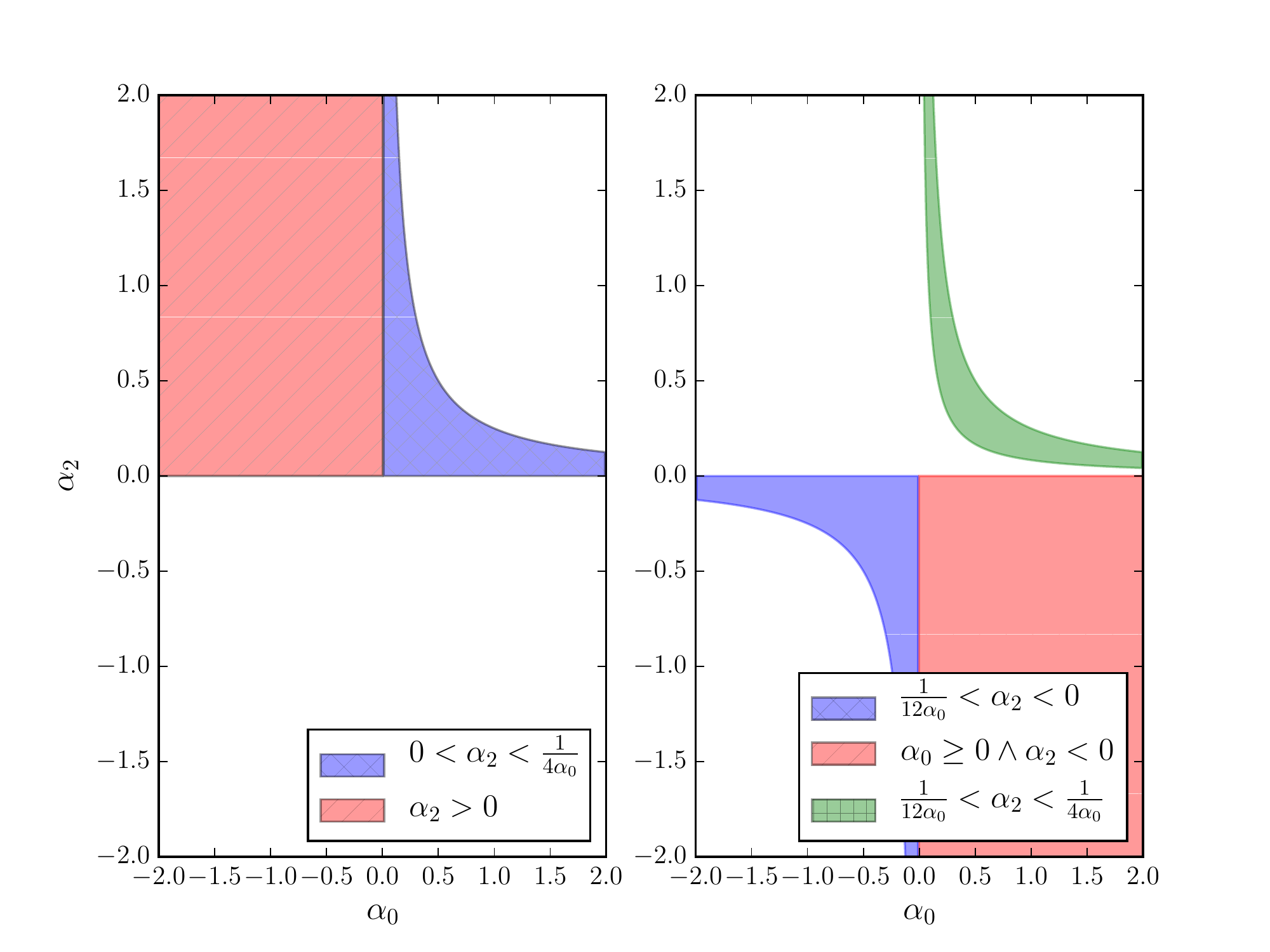}
  \caption{Allowed regions, in the $\alpha_0$--$\alpha_2$ plane, for $u_+$ and $u_-$ to be positive definite.}
  \label{cond_u_pos}
\end{figure}

In Fig.~\ref{cond_u_pos} we show the allowed region, in the $\alpha_0$--$\alpha_2$ parameter space, for $u_\pm$ to be positive. 

\begin{figure}[H]
  \includegraphics[width=.95\linewidth]{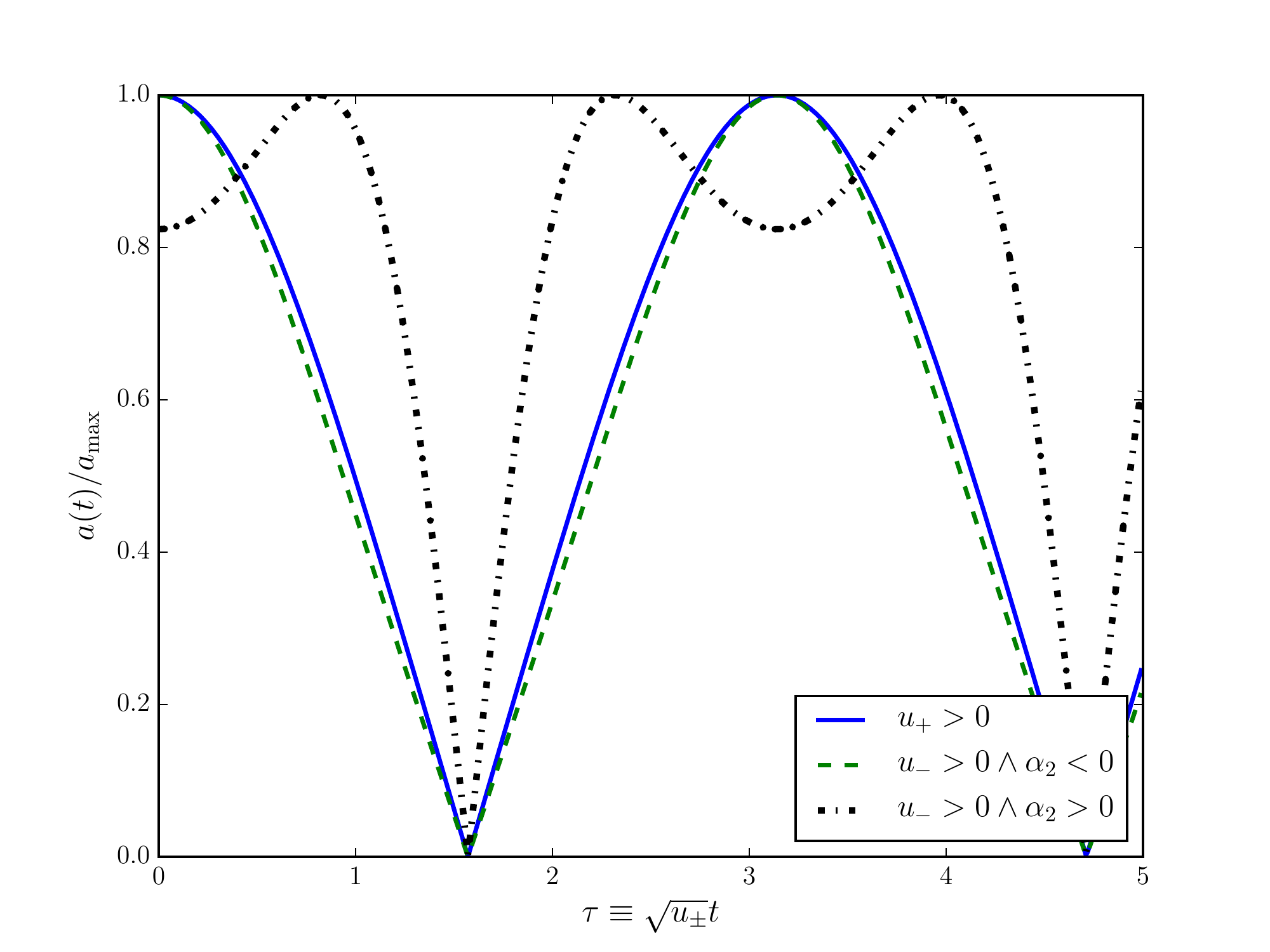}
  \caption{Behavior of the scale factor, \(a(t)\), as a function of the scaled time, $\tau$, for $u_\pm > 0$. We have normalized using the maximum value of the scale factor, $a_{\mathrm{max}}$.}
  \label{a_u_pos}
\end{figure}

\begin{figure}[H]
  \includegraphics[width=.95\linewidth]{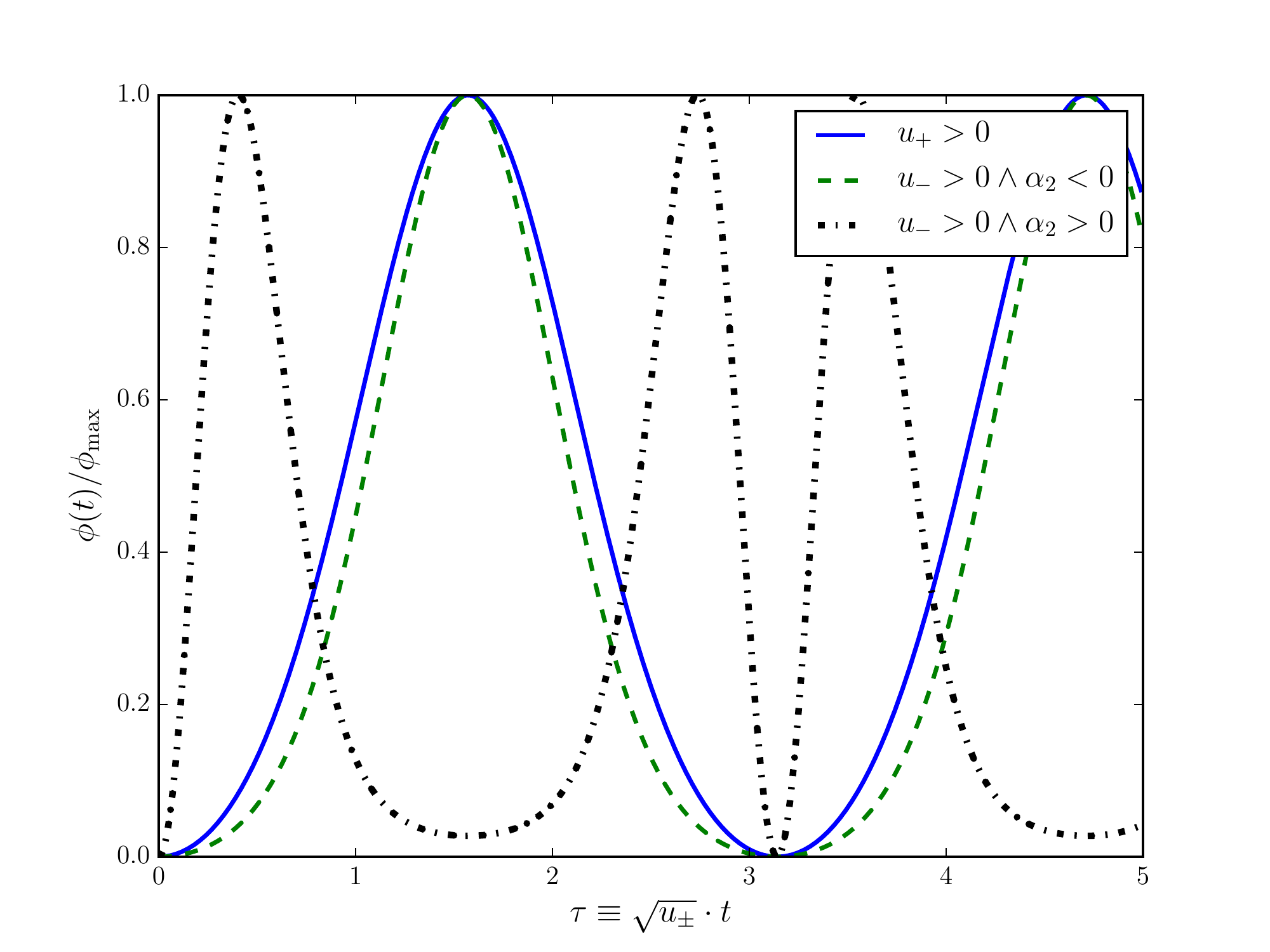}
  \caption{Behavior of the $\phi$ field as a function of the scaled time, $\tau$, for $u_\pm > 0$. We have normalized using the maximum value of the scalar field, $\phi_{\mathrm{max}}$.}
  \label{phi_u_pos}
\end{figure}

All the solutions for $u_\pm>0$ are periodic, as shown in the behavior of the scale factor in Fig.~\ref{a_u_pos}. In that case, the scale factor $a(t)$ starts from a singular point and reaches a future one after a time $t_{\rm bounce}=\pi/\sqrt{u_\pm}$. Depending on the particular region on the parameter space, the scale factor undergoes an expanding and contracting age, allowing a intermediate bounce without collapsing, before a big crunch. This is the case of the third curve in Fig.~\ref{a_u_pos}. Otherwise, it expands and collapses in a {simple} oscillatory way.
Moreover, the bouncing behavior of the extra dimension is given by the scalar field in a phase difference of $\pi/2$ with respect to $a(t)$, as shown in Fig.~\ref{phi_u_pos}. 

\subsubsection{Expanding and contracting solutions}

\begin{figure}[H]
  \includegraphics[width=.95\linewidth]{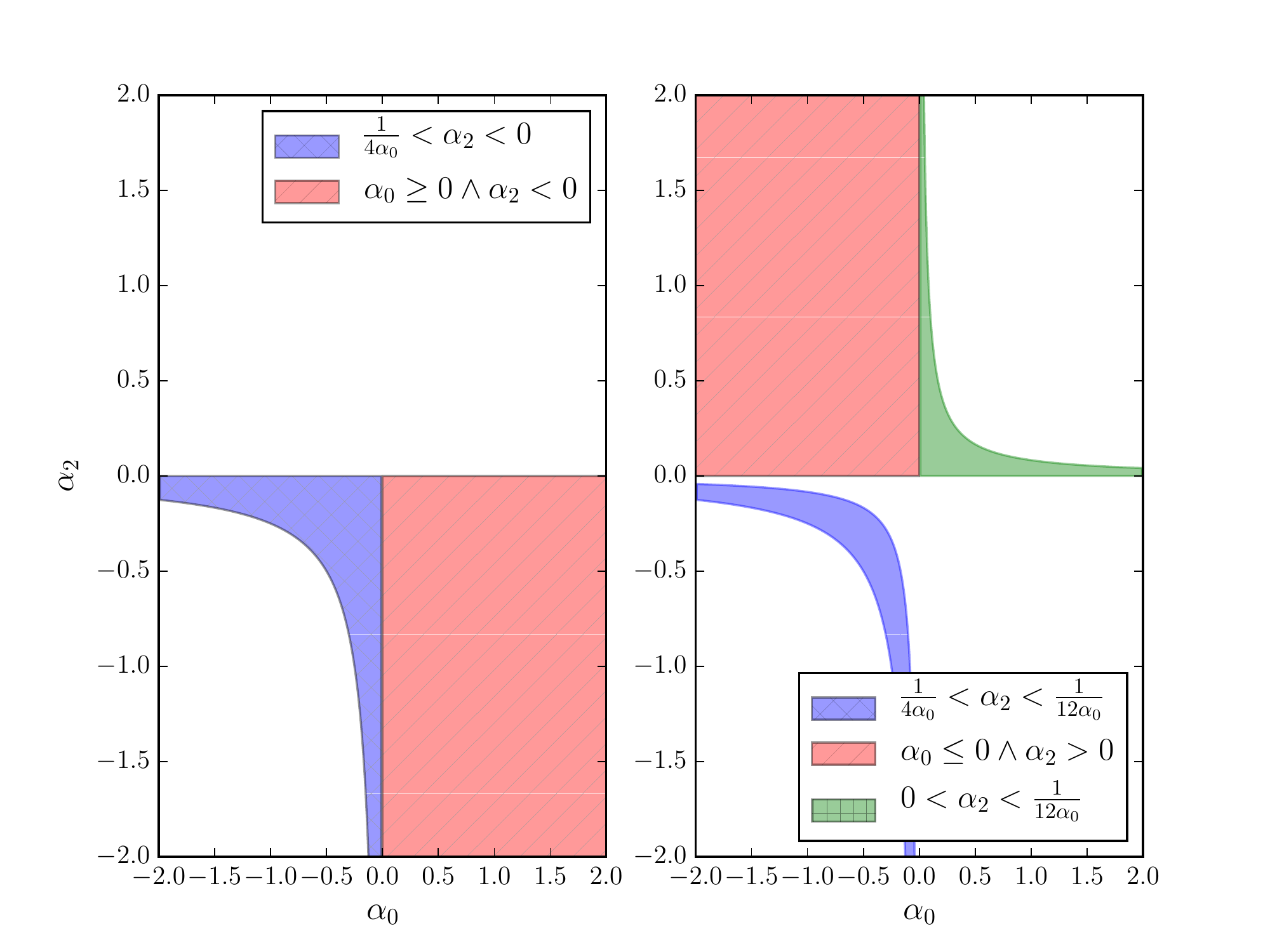}
  \caption{Allowed regions, in the $\alpha_0$--$\alpha_2$ plane, for $u_+$ and $u_-$ to be negative definite.}
  \label{cond_u_neg}  
\end{figure}

The allowed parameter space for $u_+$ and $u_-$ to be negative is shown in Fig.~\ref{cond_u_neg}, while the restriction $u_- = 0$ sets $\alpha_2 = \tfrac{1}{12 \alpha_0}$.\footnote{Notice that in our choice of parameters $u_+$ cannot vanish.}

In the regime where $u_\pm \leq 0$, there are three distinctive conducts: (i)~eternal expansion, (ii)~eternal contraction, and (iii)~initial expansion followed by an eternal contraction. In all these three scenarios, the size of the extra dimension, modulated by the $\phi(t)$ field, is reciprocal to the scale factor $a(t)$ regarding their expansive/contractive asymptotic behavior.

\begin{figure}[H]
  \includegraphics[width=\linewidth]{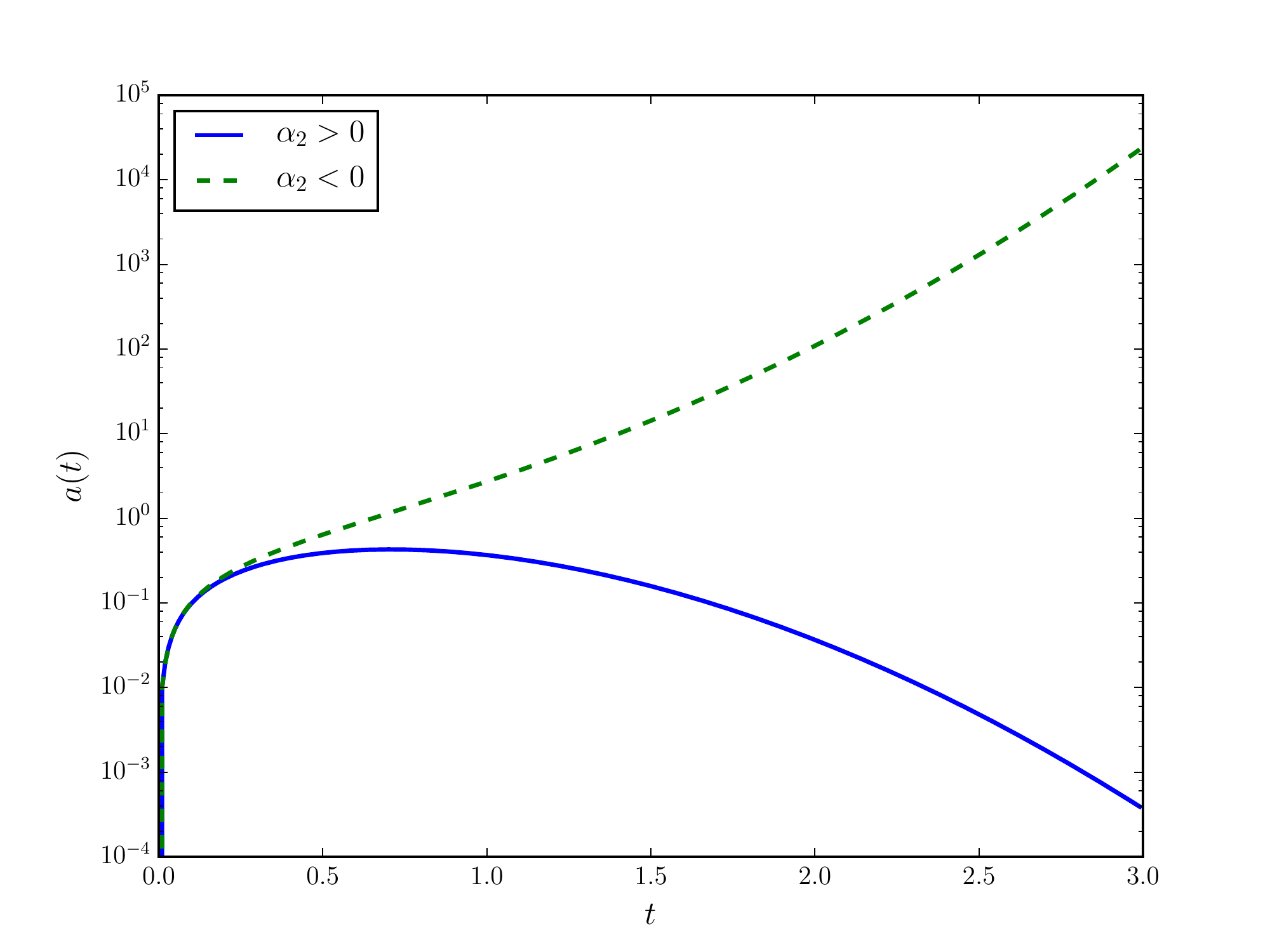} 
  \caption{Distinctive behaviors of the scale factor for $u_- = 0$.}
  \label{a_u_0}
\end{figure}

The case $u_- = 0$ has two behaviors depending on whether $\alpha_2$ is positive or negative. As shown in Fig.~\ref{a_u_0}, for positive $\alpha_2$ the solutions fit into the third category above, while for negative $\alpha_2$ the solutions expand eternally. 
\begin{figure}[H]
  \includegraphics[width=\linewidth]{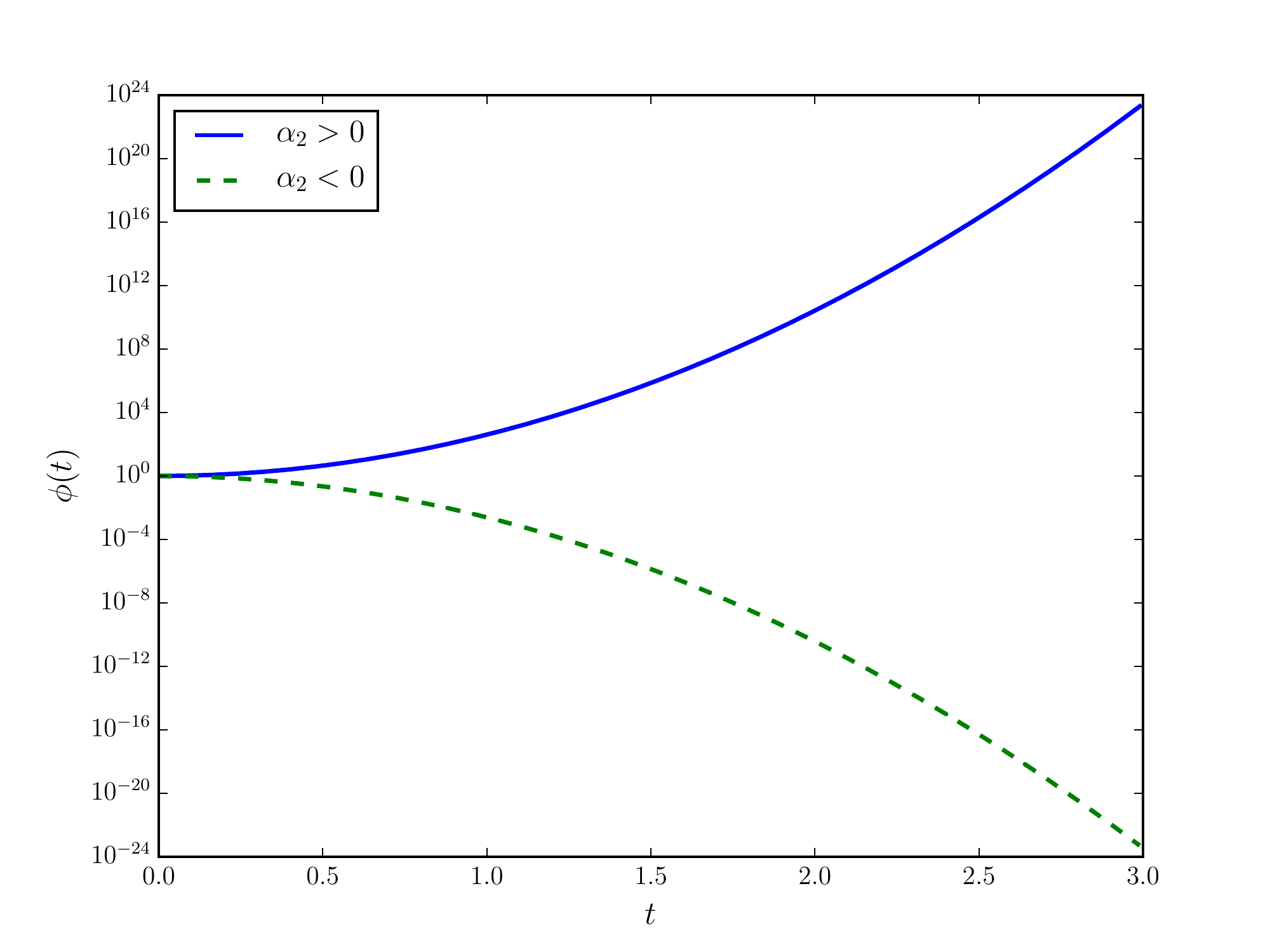}
  \caption{Distinctive behavior of the $\phi$ field for $u_- = 0$.}
  \label{phi_u_0}
\end{figure}
The $\phi$ field (see Fig.~\ref{phi_u_0}) grows infinitely for $\alpha_2 > 0$, or asymptotically goes to zero for $\alpha_2 < 0$. This latter behavior provides a {dynamical} compactification, which might serve as a mechanism to assure---at a certain time---the decoupling of the zeroth mode from the Kaluza-Klein tower.

\begin{figure}[H]
  \includegraphics[width=.95\linewidth]{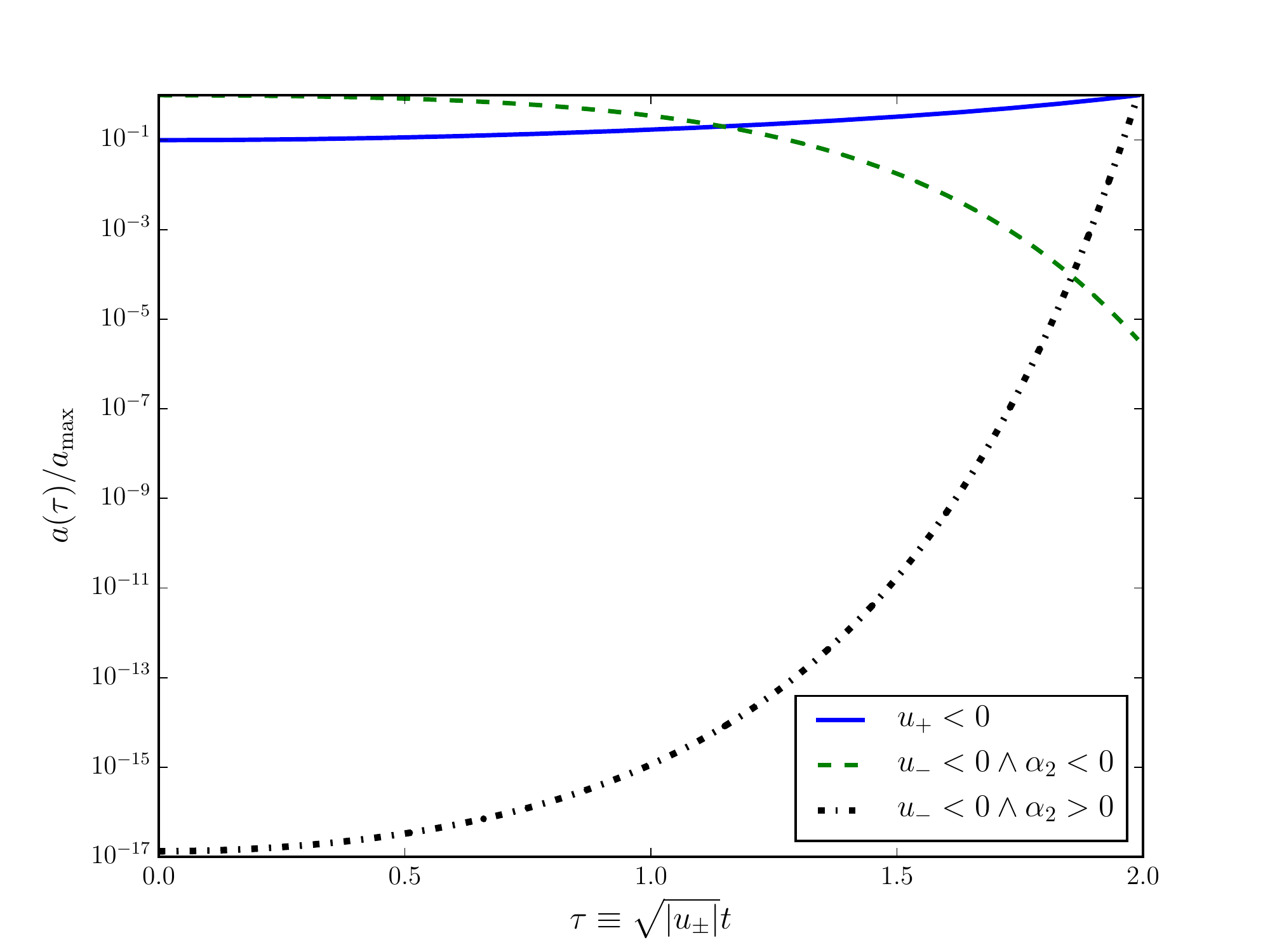}
  \caption{Different behaviors for the scale factor, as function of the scaled time $\tau$, for $u_\pm < 0$. We have normalized using the maximum value of the scale factor, $a_{\mathrm{max}}$, in the plotted region.}
  \label{a_u_neg}
\end{figure}

\begin{figure}[H]
  \includegraphics[width=.95\linewidth]{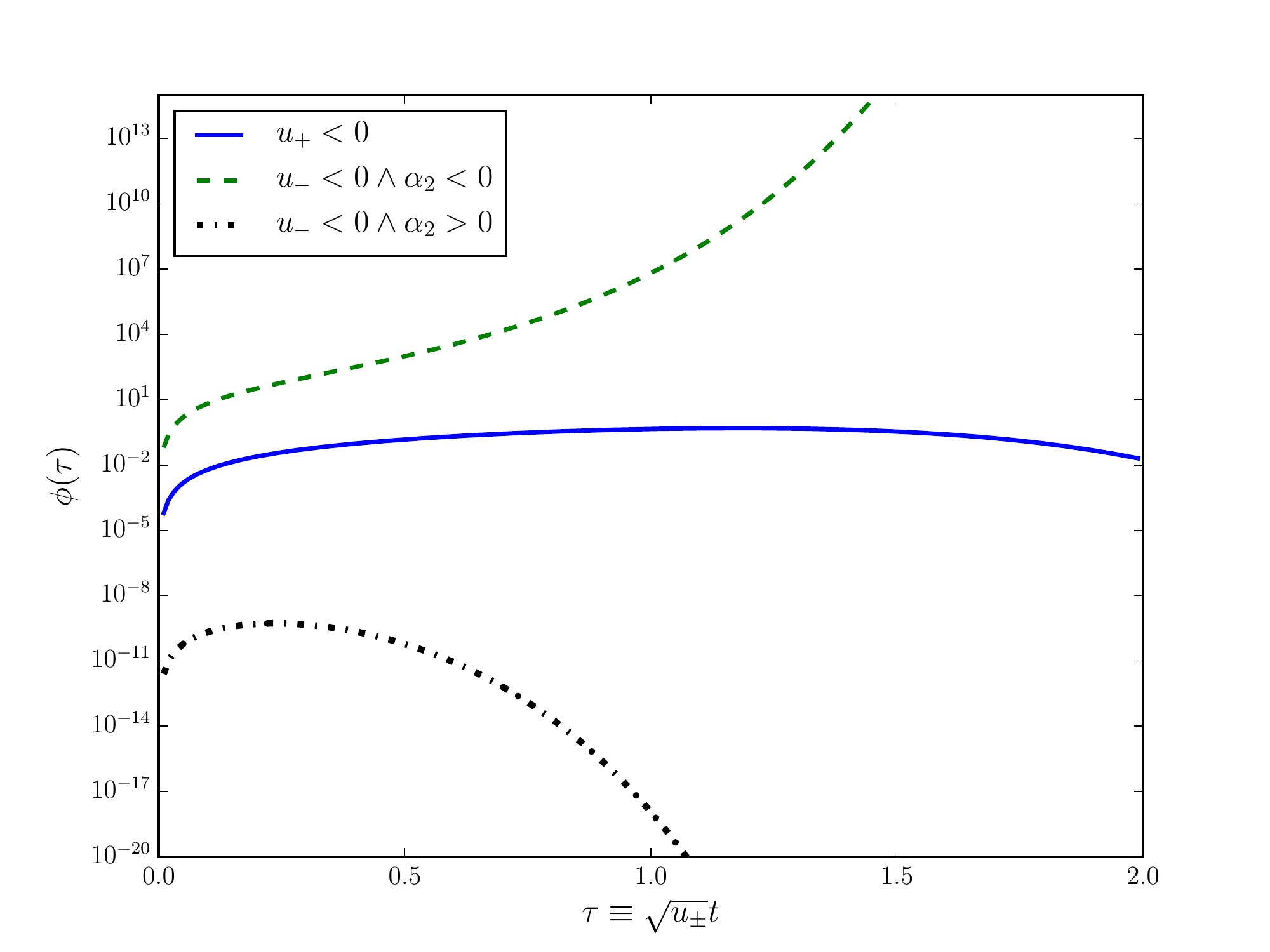}
  \caption{Different behaviors for the $\phi$ field, as function of the scaled time $\tau$, for $u_\pm < 0$. We have normalized using the maximum value of the $\phi$ field, $\phi_{\mathrm{max}}$, in the plotted region.}
  \label{phi_u_neg}
\end{figure}

The case $u_\pm < 0$ has solutions which either expand or contract eternally, as shown in Fig.~\ref{a_u_neg}. The typical evolution of the Universe with $u_+ < 0$ is to grow infinitely, while for negative $u_-$ the expansion (contraction) corresponds to $\alpha_2$ positive (negative). Figure~\ref{phi_u_neg} shows the behavior of $\phi$ for this case. 
For all these solutions the scale factor remains finite at $t = 0$, presenting no initial singularity. It has been reported that the Gauss-Bonnet term can prevent the Universe from expanding from an initial singularity~\cite{Deruelle:1986iv,Henriques:1986jw,*Ishihara:1986if}, which also applies to this case.

\subsection{Effective energy density and pressure}

Equation~\eqref{equation} can be written in a familiar form, by means of~\eqref{curvature decomp}, as
\begin{align}\label{einstein equation}
  -\frac{1}{2}\epsilon_{abcd}\left(\tilde{R}^{bc} - \frac{\Lambda}{3}\,e^b\wedge e^c\right)e^d &= \kappa_{G}\,\tau^{\mathrm{eff}}_a,
\end{align}
with $\Lambda \equiv -6\alpha_0/\alpha_1$, $\kappa_{G} \equiv 2/\alpha_1$, and 
\begin{align}
  \label{taua}
  \tau_a^{\mathrm{eff}}&= \tfrac{1}{2}\epsilon_{abcd}\Big[\tfrac{1}{2}\alpha_1\left(\kappa^b_{\ l}\wedge\kappa^{lc}+\tilde{\text{D}}\kappa^{bc}\right)\wedge e^d\\
    & \quad +\left(\text{D}\gamma^b+\tfrac{1}{2}\gamma^b\text{d}\ln\phi\right)\wedge\left(\tfrac{1}{2}\alpha_1e^c\wedge e^d+\alpha_2R^{cd}\right)\Big], \notag
\end{align}
where $\kappa^{ab}$ is the contorsion $1$-form defined as the torsional correction to the spin connection [see Eq.~\eqref{spin separation}].
The contributions of $F$, $\alpha^{ab}$, and $\beta^a$ have not been taken into account because, for the purposes of this article, they vanish. The left-hand side of Eq.~\eqref{einstein equation} is the $3$-form whose Hodge dual yields the Einstein equations with cosmological constant in four dimensions.
We identify the right-hand side as an energy-momentum $3$-form induced by the geometry, which behaves as a perfect fluid. It contains the torsional and the higher-dimensional degrees of freedom.

The energy density $\rho$ and the pressure $p$ are obtained through the identities
\begin{align}
  \tau_0^{\mathrm{eff}} &= -\frac{1}{3!} \, \rho \, \epsilon_{0ijk} \; e^i\wedge e^j\wedge e^k,\\
  \tau_i^{\mathrm{eff}} &= -\frac{1}{2} \,p \, \epsilon_{0ijk} \; e^0\wedge e^j\wedge e^k,
\end{align} 
giving
\begin{align}
  \label{rho}
  \rho &= -\frac{3}{2}\alpha_1\left(h^2-f^2+2Hh\right) \\
  & \quad + 3\omega\gamma\left[\frac{1}{2}\alpha_1+\alpha_2\left(\omega^2+\frac{k}{a^2}-f^2\right)\right], \notag
  \\
  \label{presion}
  p &= \alpha_1\left[\dot{h}+Hh+\frac{1}{2}\left(h^2-f^2+2Hh\right)\right]\\
  & \quad - 2\omega\gamma\left[\frac{1}{2}\alpha_1+\alpha_2\left(\dot{\omega}+H\omega\right)\right]\notag\\
  & \quad +\left(\dot{\gamma}+\frac{1}{2}\gamma\dot{\Phi}\right)\left[\frac{1}{2}\alpha_1+\alpha_2\left(\omega^2+\frac{k}{a^2}-f^2\right)\right]. \notag
\end{align}
Using the solutions obtained in the previous section, these expressions are
\begin{align}
  \label{our-rho}
  \rho &= \frac{3}{2}\alpha_1\left(H^2 + \frac{k}{a^2}\right) + 3\alpha_0, \\
  \label{our-p}
  p &= -\alpha_1\left(\dot{H} + \frac{3}{2}\,H^2 + \frac{k}{2a^2}\right) - 3\alpha_0,
\end{align}
and satisfy the continuity equation $\dot{\rho}+3H\left(\rho+p\right)=0$. 

For $u_\pm<0$, the energy density remains finite at the beggining. From Eqs.~\eqref{our-rho} and~\eqref{our-p} one can see that the induced energy density and pressure are not positive-definite quantities. In fact, the Universe undergoes an accelerating expansion phase due to the presence of torsion and the extradimensional fields. 


\section{Discussion and Conclusions\label{conclusions}}

In this work we have presented the dimensional reduction of the five-dimensional Lovelock-Cartan theory introduced in Ref.~\cite{Mardones:1990qc}, under the assumption that the compact dimension has $S^1$ topology. An interesting feature of these theories is that, unlike the Einstein-Cartan theory, torsion is allowed to propagate.

Although there are several generalizations of gravitational models which include higher-order terms in curvature and torsion---in either four or higher
dimensions~\cite{Puetzfeld:2004yg,Baekler:2011jt,Fabbri:2012qr}--- we highlight that in our model the presence of higher-order terms in torsion are introduced through the Gauss-Bonnet term, accompanied by a sole extra coupling constant.
Moreover, Lovelock-Cartan gravity ensures that the field equations are first-order equations (since the Lagrangian can be written without the use of the Hodge dual), which is not the case in most gravitational models with higher order-curvature and torsion.

In the generic reduction, the effective theory has a spin-2 particle, a spin-1 $U(1)$ gauge boson, and a spin-0 scalar particle coming from the decomposition of the metric, while the decomposition of the spin connection introduces three extra fields: two spin-1 particles, $\alpha^{ab}$ and $\gamma^a$, and the 1-form $\beta^a$ yields a new spin-2 particle, a spin-1 particle, and a spin-0 particle. These new fields do not transform under the gauge group (see Sec.~\ref{sec:bianchi}). 

We used the most general ansatz compatible with the cosmological symmetries (see Appendix~\ref{homotropic}) to find solutions of the Friedmann-Robertson-Walker class in the four-dimensional theory. Though we restrict ourselves to the class of theories with $\Delta > 0$, in Ref.~\cite{Canfora:2014iga} the case of negative $\Delta$ is considered (without torsion). The field equations for the cosmological ansatz ensure that the solutions with nontrivial torsion have vanishing $\alpha^{ab}$ and $\beta^a$. The solutions are parametrized by $u_\pm$, related to the fundamental couplings of the theory through Eq.~\eqref{u}, and exhibit three different sectors depending on whether this parameter is positive, negative, or zero. The behavior of the universes described by our solutions is fourfold: (i)~eternal expansion, (ii)~eternal contraction, (iii)~initial expansion with asymptotic contraction, and (iv)~bouncing.

Among the different cases, there are some solutions that do not start from an initial singularity; however, this is not the generic behavior. Moreover, all of our cosmological solutions are free of the future singularities categorized in Ref.~\cite{Shojai:2015dpa}. In addition, the size of the extra dimension is driven by the scalar field $\phi(t)$ and its behavior at late times is compatible with a dynamical mechanism for compactification, at certain regions of the parameter space (see Figs.~\ref{phi_u_0} and~\ref{phi_u_neg}). Another interesting feature is that in all of our cosmological scenarios, the scale factor does not depend on the $k$ parameter which modulates the metric structure.

The field equation~\eqref{equation} admits the interpretation of the Einstein equations, where the energy-momentum $3$-form is induced by the torsion and the extradimensional degrees of freedom. It is important to mention that the induced energy-momentum form is nonstandard, in the sense that it inherits a coupling between matter and curvature [see Eq.~\eqref{taua} for the definition]. Moreover, the energy density and the pressure are not positive definite, 
providing a playground for a gravitational explanation for the accelerated expansion of the Universe.



As we can see from Eq.~\eqref{homotorsion}, $f$ corresponds to the completely antisymmetric part of torsion. Its solution given in Eq.~\eqref{f2sol} shows that it can be imaginary for certain periods of time. Additionally, it is the only function sensitive to the value of $k$. Thus, if a reality condition is imposed over $f$ for certain values of the coupling constants, it would provide a distinction criteria among open, flat, and closed universes. In the light of this, and given the experimental evidence supporting the flatness of the Universe (see Ref.~\cite{Spergel:2006hy}), we restrict the discussion to the case where $k=0$.\footnote{The cases where $k = \pm1$ can be studied analogously.} A reality condition over $f$ in Eq.~\eqref{f2sol} is satisfied within the region 
\begin{align*}
 & & 3\alpha_1 \pm \sqrt{6\Delta} &\geq 0 & &\mbox{and}& \alpha_2 &>0, \\
 & \mbox{or,} & 3\alpha_1 \pm \sqrt{6\Delta} &\leq 0 & &\mbox{and}& \alpha_2&<0.
\end{align*}
This should be taken into account to restrict the parameter region studied in Sec.~\ref{sec:sol}.  
A similar behavior for this function was found in Ref.~\cite{Toloza:2013wi}. Notice that the antisymmetric part of torsion is unseen by classical particles following geodesics and it does not couple minimally to spin-0 or spin-1 bosons. However, it appears as an effective interaction term for the Dirac Lagrangian in spacetimes with torsion~\cite{Hehl:1976kj}
\begin{align*}
  \mathcal{L}_{\text{Int}}&=-\frac{i}{8}T_{\alpha\mu\nu}\bar{\Psi}\Gamma^{\alpha\mu\nu}\Psi =\frac{3}{2}f\bar{\Psi}\Gamma^0\Gamma^5\Psi,
\end{align*}
which couples to the fermionic axial current. For imaginary values of $f(t)$, the authors in Ref.~\cite{Toloza:2013wi} argued that the loss of unitarity and the violation of the current conservation can be interpreted as particle creation. 

The vacuum cosmological solutions presented here provide a useful arena to isolate the effects of the induced fields and probe their consequences in physically important scenarios. In that sense, and since they are absent in the cosmological ansatz, a four-dimensional isotropic configuration might contribute to interpret the $\alpha^{ab}$ and $\beta^a$ fields as gravitational hairs in black-hole solutions, or wormhole-supporting matter. 

The gravitational principles we take for granted can be relaxed to construct a logically consistent theory. This approach might provide an explanation for the phenomena at the limit of the current understanding of the Universe, such as its content and evolution. This work shows that the renouncing of the metric description of gravity, attached to a higher-dimensional spacetime, can radically affect the way we perceive the gravitational degrees of freedom and, therefore, what our measurements read as the energy content of the Universe. 





\begin{acknowledgments}
  The authors would like to thank A.~R.~Zerwekh, I.~Schmidt,
  S.~Kovalenko, N.~A.~Neill, C.~O.~Dib, O.~Miskovic, R.~Gannouji,
  A.~Toloza, A.~Cisterna, and J.~Zanelli for valuable comments on this
  work. \mbox{O.~C.-F.} has been partially supported by the PAI
  Project No.~79140040 (CONICYT--Chile) and F.~R. by the Fondecyt
  Project No.~1160423. The Centro Cient\'ifico Tecnol\'ogico de
  Valpara\'iso (CCTVal) is funded by the Chilean Government through
  the Centers of Excellence Basal Financing Program FB0821 of CONICYT.
\end{acknowledgments}

\appendix

\section{Riemann-Cartan Geometry\label{Riemann-Cartan}}

In Riemann-Cartan geometry, both the vielbein $e^A$ and the spin connection $\omega^{AB}$ are independent features of the manifold.\footnote{Here, we will drop the hats since we will not deal with the KK decomposition.} The spin connection can be decomposed in a Riemannian part, which is metric dependent, $\tilde{\omega}^{AB}$ satisfying $\tilde{\mbox{D}}e^A=0$, and a contorsion piece $\kappa^{AB}=-\kappa^{BA}$, such that 
\begin{equation}\label{spin separation}
  \omega^{AB}=\tilde{\omega}^{AB}+\kappa^{AB}.
\end{equation} 
Therefore, the torsion 2-form defined as the covariant derivative of the vielbein with respect to the total spin connection, is
\begin{equation}
  T^A=\kappa^A_{\ B}\wedge e^B.
\end{equation}
On the other hand, the curvature $2$-form also suffers corrections with respect to the Riemannian one, due to the presence of torsional degrees of freedom. This can be seen explicitly by taking the definition of curvature in Eq.~\eqref{curvadef} and~\eqref{spin separation} to find
\begin{equation}\label{curvature decomp}
  R^{AB} = \tilde{R}^{AB} + \tilde{\mbox{D}}\kappa^{AB} + \kappa^A_{\ C}\wedge\kappa^{CB},
\end{equation}
where $\tilde{R}^{AB}$ is the Riemannian curvature constructed with $\tilde{\omega}^{AB}$.

\section{Isotropic-Homogeneus Ansatz\label{homotropic}}

The cosmological principle demands the spatial section of spacetime to be isotropic and homogeneous. This means that the fields involved in the model must be compatible with this assumption. A spacetime is isotropic with respect to certain point $P$ if after a rotation with respect to an axis passing through $P$, all the geometrical properties remain invariant. Thus, the spacetime looks the same in all directions. Homogeneity is understood such that the spacetime looks the same from every point $P$. These two assumptions are translated in the Killing equations, which are the vanishing of the Lie derivatives of the fields along the vectors which generate the symmetries $\{\zeta^\lambda_{i}\}$. 

The set of Killing vectors $\{\zeta^\lambda_i\}$ are the generators of $SO(3)$, which generate the spatial rotations in three dimensions, $\mathcal{J}_i=\epsilon_{ijk}x_j\partial_k$, and the Killing vectors associated with spatial translations $\mathcal{P}_i=\sqrt{1-kr^2}\partial_i$, satisfying the algebra
\begin{align*}
  \left[\mathcal{J}_i,\mathcal{J}_j\right]&=\epsilon_{ijk}\mathcal{J}_k,\\
  \left[\mathcal{P}_i,\mathcal{P}_j\right]&=-k\epsilon_{ijk}\mathcal{J}_k,\\
  \left[\mathcal{J}_i,\mathcal{P}_j\right]&=\epsilon_{ijk}\mathcal{P}_k.
\end{align*}

In particular the Killing equations for the metric tensor and the torsion tensor are
\begin{align}
  \label{killing metric}
  \text{\textsterling}_i g_{\mu\nu}&=\zeta^\lambda_i\partial_\lambda g_{\mu\nu} + g_{\lambda\nu}\partial_\mu\zeta^\lambda_i  + g_{\mu\lambda}\partial_\nu\zeta^\lambda_i = 0,
  \\
  \text{\textsterling}_i T^\alpha_{\ \mu\nu}&=\zeta^\lambda_i\partial_\lambda T^\alpha_{\ \mu\nu} - T^\lambda_{\ \mu\nu} \partial_\lambda \zeta^\alpha_i + T^\alpha_{\ \lambda\nu} \partial_\mu\zeta^\lambda_i  \notag \\
  &\quad + T^\alpha_{\ \mu\lambda} \partial_\nu\zeta^\lambda_i = 0,
\end{align}
where $T^\alpha_{\ \mu\nu}$ are the components of the torsion 2-form defined by $T^a=\frac{1}{2}e^a_{\ \alpha}T^\alpha_{\ \mu\nu}\mbox{d}x^\mu\wedge\mbox{d}x^\nu$. The same must hold for the new fields
\begin{align}
  \label{killing vector}
  \text{\textsterling}_i A_\mu&=\zeta^\lambda_i\partial_\lambda A_\mu + A_\lambda \partial_\mu\zeta^\lambda_i = 0,\\
  \text{\textsterling}_i \phi &=\zeta^\lambda_i\partial_\lambda\phi=0,
\end{align}
and also for the components of $\alpha_{\mu\nu}=-\alpha_{\nu\mu}$, $\beta_{\mu\nu}$ and $\gamma_\mu$, defined such that
\begin{align*}
  \alpha^{ab}&=E^{a\mu}E^{b\nu}\alpha_{\mu\nu},\\
  \beta^a&=E^{a\mu}\beta_{\mu\nu}\mbox{d}x^\nu,\\
  \gamma^a&=E^{a\mu}\gamma_\mu,
\end{align*} 
and whose Killing equations are analogous to the tensorial, Eq.~\eqref{killing metric}, and vectorial one, Eq.~\eqref{killing vector}. These requirements on the fields are translated to a set of first-order differential equations whose most general solution determines our ansatz structure for Eqs.~\eqref{vielbein cosmo}--\eqref{gamma cosmo}. 

The components of the Lorentz curvature and torsion for the isotropic-homogeneus ansatz are 
\begin{align*}
  R^{0i}&=\left(\dot{\omega}+H\omega\right) \, e^0\wedge e^i+f\omega\epsilon^i{}_{jk} \, e^j\wedge e^k,\\
  R^{ij}&=\left(\omega^2+\tfrac{k}{a^2}-f^2\right) \, e^i\wedge e^j
  -\left(\dot{f}+Hf\right)\epsilon^{ij}{}_{k} \, e^0\wedge e^k,
\end{align*}
and 
\begin{equation}\label{homotorsion}
  T^0=0,\quad T^i=-h \, e^0\wedge e^i+f\epsilon^i{}_{jk} \, e^j\wedge e^k,
\end{equation}
respectively. The functions $a(t)$, $\omega(t)$, $f(t)$, $h(t)$, and $H(t)$ were defined in Sec.~\ref{cosmos}.

\section{Time dependence on the solutions\label{solutions t}}

We supply here the time-dependent expressions for the cosmological solutions, given the different values of $u_\pm$, in terms of the dimensionless parameter \mbox{$\tau=\sqrt{|u_\pm|}(t-t_0)$}.
\vspace{.1cm}

For $\bm{u_\pm>0}$,
\begin{align}
    \gamma(t)&=-\sqrt{u_\pm}\cot\tau,\\
    \phi(t)&=\phi_0\sin^2\tau\exp\left[\tfrac{\mp\sqrt{6\Delta}}{4\alpha_2 u_\pm}\cos^2\tau\right],\\
    a(t)&=\tfrac{a_0}{\sqrt{u_\pm}}|\cos\tau|\exp\left[\tfrac{\pm\sqrt{6\Delta}}{24\alpha_2 u_\pm}\cos^2\tau\right],\\
    f^2(t)&=u_\pm\tfrac{k}{a_0^2}\sec^2\tau\exp\left[\tfrac{\mp\sqrt{6\Delta}}{12\alpha_2u_\pm}\cos^2\tau\right]\notag\\
    &\quad +u_\pm\tan^2\tau + \tfrac{3\alpha_1\pm\sqrt{6\Delta}}{6\alpha_2}.
  \end{align}

For $\bm{u_\pm=0}$,
  \begin{align}
    \gamma(t)&=0,\\
    \phi(t)&=\phi_0\exp\left[\tfrac{\mp\sqrt{6\Delta}}{4\alpha_2}(t-t_0)^2\right],\\
    a(t)&=a_0|t-t_0|\exp\left[\tfrac{\pm\sqrt{6\Delta}} {24\alpha_2}(t-t_0)^2\right],\\
    f^2(t)&=\tfrac{k}{a_0^2(t-t_0)^2}\exp\left[\tfrac{\mp\sqrt{6\Delta}}{12\alpha_2}(t-t_0)^2\right]\notag\\
    &\quad +(t-t_0)^{-2} + \tfrac{3\alpha_1\pm\sqrt{6\Delta}}{6\alpha_2}.
  \end{align}

For $\bm{u_\pm<0}$,
  \begin{align}
    \gamma(t)&=-\sqrt{-u_\pm}\coth\tau,\\
    \phi(t)&=\phi_0\sinh^2\tau\exp\left[\tfrac{\mp\sqrt{6\Delta}}{4\alpha_2u_\pm}\cosh^2\tau\right],\\
    a(t)&=\tfrac{a_0}{\sqrt{-u_\pm}}\cosh\tau\exp\left[\tfrac{\pm\sqrt{6\Delta}} {24\alpha_2u_\pm}\cosh^2\tau\right],\\
    f^2(t)&=-u_\pm\tfrac{k}{a_0^2}\cosh^{-2}\tau\exp\left[\tfrac{\mp\sqrt{6\Delta}}{12\alpha_2u_\pm}\cosh^2\tau\right]\notag\\
    &\quad -u_\pm\tanh^2\tau + \tfrac{3\alpha_1\pm\sqrt{6\Delta}}{6\alpha_2}.
  \end{align}
%

\end{document}